%% file: main.tex
\pdfoutput=1
\documentclass[journal]{IEEEtran}
\usepackage{amsmath,amsfonts}
\usepackage{algorithm}
\usepackage{algpseudocode}
\usepackage{algorithmicx}
\usepackage{array}
\usepackage[caption=false,font=normalsize,labelfont=sf,textfont=sf]{subfig}
\usepackage{textcomp}
\usepackage{stfloats}
\usepackage{url}
\usepackage{verbatim}
\usepackage{graphicx}
\usepackage{cite}
\usepackage{xcolor}
\usepackage{multirow}
\usepackage{footnotehyper}
\usepackage{threeparttable}
\makesavenoteenv{tabular}
\begin{document}
\title{A Heterogeneous In-Memory Computing Cluster For Flexible End-to-End Inference of Real-World Deep Neural Networks}

\author{\IEEEauthorblockN{Angelo Garofalo\IEEEauthorrefmark{1}, Gianmarco Ottavi\IEEEauthorrefmark{1}, Francesco Conti\IEEEauthorrefmark{1},\\ Geethan Karunaratne\IEEEauthorrefmark{2}\IEEEauthorrefmark{3}, Irem Boybat\IEEEauthorrefmark{2}, Luca Benini\IEEEauthorrefmark{3}\IEEEauthorrefmark{1}, Davide Rossi\IEEEauthorrefmark{1} 
\thanks{This work was supported by the EU Horizon 2020 Research and Innovation projects WiPLASH (g.a. no. 863337) and European Pilot (g.a. 101034126).}} \\
\IEEEauthorblockA{{\small\textit{\IEEEauthorrefmark{1}University of Bologna, Bologna, Italy} \quad \textit{\IEEEauthorrefmark{2}IBM Research Europe, Zurich, Switzerland} \quad \textit{\IEEEauthorrefmark{3}ETH Zurich, Zurich, Switzerland} \\
$\{$angelo.garofalo, gianmarco.ottavi2, davide.rossi, f.conti$\}$@unibo.it \quad
$\{$lbenini$\}$@iis.ee.ethz.ch \quad
$\{$kar, ibo$\}$@zurich.ibm.com}}}



\maketitle

\begin{abstract}

Deployment of modern TinyML tasks on small battery-constrained IoT devices requires high computational energy efficiency. 
Analog In-Memory Computing (IMC) using non-volatile memory (NVM) promises major efficiency improvements in deep neural network (DNN) inference and serves as on-chip memory storage for DNN weights.
However, IMC's functional flexibility limitations and their impact on performance, energy, and area efficiency are not yet fully understood at the system level.
To target practical end-to-end IoT applications, IMC arrays must be enclosed in heterogeneous programmable systems, introducing new system-level challenges which we aim at addressing in this work. 
We present a heterogeneous tightly-coupled clustered architecture integrating 8 RISC-V cores, an in-memory computing accelerator (IMA), and digital accelerators.
We benchmark the system on a highly heterogeneous workload such as the Bottleneck layer from a MobileNetV2, showing 11.5$\times$ performance and 9.5$\times$ energy efficiency improvements, compared to highly optimized parallel execution on the cores.
%
Furthermore, we explore the requirements for end-to-end inference of a full mobile-grade DNN (MobileNetV2) in terms of IMC array resources, by scaling up our heterogeneous architecture to a multi-array accelerator.
Our results show that our solution, on the end-to-end inference of the MobileNetV2, is one order of magnitude better in terms of execution latency than existing programmable architectures and two orders of magnitude better than state-of-the-art heterogeneous solutions integrating in-memory computing analog cores.

\end{abstract}

\begin{IEEEkeywords}
In-memory computing, RISC-V, Heterogeneous computing architecture, MobileNetV2
\end{IEEEkeywords}

\input{text/01_introduction_v2.tex}
\input{text/02_related_work_v2.tex}
\input{text/03_background_v2.tex}
\input{text/04_body_v3.tex}

\input{text/05_results_and_discussion_v2.tex}

\input{text/06_MobileNetV2_inference}
\input{text/07_SOA_Comparison_v2.tex}

\input{text/08_conclusion.tex}

\bibliographystyle{IEEEtran}
\bibliography{IEEEfull, pulp_ima.bib}

\input{bio_text/authors_biography}

\end{document}

%% file: text/01_introduction_v2.tex
\section{Introduction}
Analog in-memory computing (AIMC) performs data processing in situ within memory arrays. Matrix-vector multiplication (MVM) operands can be mapped on the cross-bars of a non-volatile (NV) memory array and the \textit{dot product} operation is performed entirely in the analog domain, making IMC devices promising candidates to accelerate DNN workloads, and overcome the well-known memory bottleneck affecting traditional AI digital accelerators~\cite{InMComp}. Both charge-based memory technologies (e.g. SRAM, DRAM, and flash) and resistance-based memory technologies (e.g. RRAM, PCM, and STT-MRAM) can serve as elements for such computational units~\cite{sebastian2020memory}.

Several demonstrations of AIMC-based architectures have appeared in the field of Deep Neural Network (DNN) inference acceleration, showing outstanding peak energy efficiency in the order of hundreds of TOPS/W~\cite{InMComp, sebastian2020memory}. Industry interest in this technology is growing~\cite{axelera, mythicHC2018}.
From a research perspective, several prototypes claimed tens to hundreds of TOPS/W by exploiting many different approaches, with a quite diverse set of choices in terms of numerical precision and underlying memory technologies~\cite{InMComp, sebastian2020memory}.

However, several fundamental challenges are still open to achieve the claimed levels of efficiency at full-application scale: the intrinsic variability of analog computing both in the charge-based and resistive domain~\cite{sebastian2020memory}; difficulties in dealing with low-precision computations that are often the only ones supported by AIMC-based architectures~\cite{sebastian2020memory}; the necessity of specialized training~\cite{seb2020mixed}. Most prominently, a key issue is the limited flexibility of IMC arrays, which 
%
are extremely efficient on MVM or similar vector operations, but they are not flexible enough to sustain other types of workloads. To tackle this limitation, a prominent solution is to couple either general-purpose processors~\cite{jia2020programmable} or specialized digital accelerators~\cite{zhou2021analognets} with analog in-memory computing cores. This allows extending the functionality of In-Memory Accelerators (IMA), creating heterogeneous analog/digital computing fabrics, connected to the system bus~\cite{jia2020programmable}. However, this integration poses severe concerns at the system level, mainly on two aspects: bandwidth and flexibility. 

First, IMC acceleration moves the challenge towards ensuring efficient data movement within the system. In the case of volatile technologies, such as SRAM-based IMC, the weights of the DNN must be stored in non-volatile memory (external or internal to the system). This requires additional energy and time to move the data that must be stored into the cells of the IMA, anytime the cross-bar is programmed~\cite{InMComp}. When considering non-volatile technologies, such as Flash, ReRAM or PCM-based IMC arrays, weights are directly stored into the cross-bar, with no need for marshaling operations. However, previous concerns continue to affect the activations that must be moved at the boundaries of the IMC array, to perform MVMs. Taking this into account, efficient integration of IMC into heterogeneous systems requires an optimized interface design between the highly parallel IMC inputs/outputs, the programmable cores, and the rest of the system: low bandwidth and high communication latency between the processor and the IMA might create a major bottleneck~\cite{jia2020programmable}.

Second, as a consequence of Amdahl's effect, accelerating MVM operators with an IMA moves the performance bottleneck on all the other computation needed to accomplish a certain task, which must be performed on the digital part of the system. Complex real-world neural networks mix MVMs with other workloads such as residuals, activation functions, or depth-wise convolutions; coupling the IMA with a single core, as has recently been proposed~\cite{jia2020programmable}, will likely hit Amdahl's effect caused by the single-core bottleneck, hindering the whole computation performance.

In this work, we address the system-level challenges of analog IMC by exploiting extreme heterogeneity. 
%
The main contributions of this paper are the following:
\begin{itemize}
    \item We design a heterogeneous tightly-coupled shared-memory cluster that integrates 8 fully programmable RISC-V processors, an analog in-memory computing accelerator (IMA), and a dedicated digital block to accelerate depth-wise convolutions; we present a post place\&route silicon-ready implementation targeting GlobalFoundries 22nm FDX technology;
    \item We optimize the interfaces between the analog IMA and the rest of the system to match the computing and IO requirements of the IMA, achieving performance as high as 958 GOPS on MVMs, more than 90\% of its peak theoretical throughput, surpassing by one order of magnitude other approaches where the IMA is connected through a low-bandwidth, high-latency system bus~\cite{jia2020programmable}; 
    \item We benchmark the system on a bottleneck layer, representative of modern DNNs exploiting heavily heterogeneous layers such as point-wise, depth-wise convolutions and residuals, in terms of performance and energy efficiency. The analog/digital synergistic approach demonstrates full mitigation of Amdahl's effect, showing $2.6\times$ better performance and $2.8\times$ better energy efficiency compared to executing the layers on our previous work that integrates only 8 programmable cores and the IMC analog array~\cite{ottavi2021end};
    \item We scale up the proposed system to analyze the challenges and the hardware resources necessary to enable end-to-end inference of a MobileNetV2. Our architectural paradigm executes inference in 10$ms$ with an energy of 482$\mu J$, improving upon fully digital state-of-the-art solutions (SoA)~\cite{rossi2021vega} by $10\times$ in latency, reducing the energy consumption by $2.5\times$. Compared to SoA analog/digital architectures~\cite{jia2020programmable}, our solution shows two orders of magnitude improvements in terms of execution latency.
\end{itemize}

The manuscript is organized as follows: in Sec.~\ref{sec:Related_Work} we review the state-of-the-art; in Sec.~\ref{sec:Background} we outline the background, and in Sec.~\ref{sec:BODY} we present the heterogeneous cluster architecture. Then, in Sec.~\ref{sec:Experimental_Results} and Sec.~\ref{sec:discussion} we report the experimental results, discussions, and explorations. Finally in Sec.~\ref{sec:comparison_with_soa} we compare with state-of-the-art solutions.
Sec.~\ref{sec:conclusion} concludes the paper.

%% file: text/02_related_work_v2.tex
\section{Related Work}
\label{sec:Related_Work}
Charge-based memory technologies (e.g. SRAM~\cite{biswas2018conv}, DRAM, Flash) and non-volatile (NV) resistive memory technologies~\cite{ielmini2018memory} (e.g. ReRAM~\cite{8310400} PCM~\cite{sebastian2020memory} and MRAM~\cite{doevenspeck2020sot}) both serve as computing substrates for analog in-memory computing. In this section, we review the State-of-the-Art (SoA) advancements in in-memory computing technology, circuits, and systems. 

%
%
\subsubsection{IMC Arithmetic}
Low-bit-width integer computation is widely adopted in edge Artificial Intelligence (AI) applications, because of its higher efficiency and lower hardware cost than floating-point. In the IMC domain, the advantages are even more evident. Low bit-width data representation results in less area and power costs to design analog to digital (ADCs) and digital to analog (DACs) converters, which are predominant in IMC arrays~\cite{InMComp, sebastian2020memory}. The adoption of heavily quantized integer arithmetic (8-bit or less), especially for DNNs, is fully justified by the fact that Quantized Neural Networks (QNNs) show a negligible drop-in Top-1 accuracy compared to the full floating-point precision model, on many AI-enhanced edge applications~\cite{jacob2018quantization}. Also, noise-robust networks are an active research field for IMC deployment~\cite{joshi2020accurate}.

%

\subsubsection{SRAM technology}
The SRAM technology is the most mature one, optimized for decades to be used as volatile memory storage for digital computing architectures.
%
%
SRAMs are used to perform MVM operations both in the digital and analog domains. In the digital domain, the computation is performed coupling SRAM cells with additional near-memory logic, such as elementary gates, full adders, or adder trees, building up a digital accelerator~\cite{chih202189tops}. In the analog domain, SRAMs can map MVMs by exploiting capacitive charge redistribution mechanisms along the bit-lines of the memory array~\cite{sebastian2020memory}. Compared to the analog approach, SRAM-based digital IMC provides higher robustness to noise and process, voltage, and temperature (PVT) variations, but significantly less advantages in terms of energy efficiency~\cite{ kim20191}. 

Most SoA academic SRAM-based IMC arrays operate in the analog domain~\cite{mittal2021survey}. One of the first prototypes appeared in 2018~\cite{biswas2018conv}, targeting binary-weight Neural Networks and demonstrating top-1 accuracy comparable with software accuracy on the MNIST dataset ($\sim$98\%). 
SRAM-based IMC has been demonstrated for binary/ternary DNNs achieving 403 TOPS/W and software accuracy on ternary networks trained on the CIFAR-10 dataset~\cite{yin2020xnor}, as well as for reconfigurable bit-precision MVM operations showing80 TOPS/W~\cite{yue20212}. 
Other SRAM-based IMC architectures have been proposed, achieving similar accuracy and efficiency~\cite{lee2021fully}.
The major challenge at the circuits level, which is actively being investigated in the literature, remains the computation noise that limits the signal-to-noise ration, mainly due to the sensitivity to PVT variations~\cite{sebastian2020memory}, and non-linearities~\cite{InMComp}.

\subsubsection{Resistive Memory technology}
A new generation of IMC accelerators targets emerging resistive memory technology, driven by the much higher density scaling factor that these technologies offer compared to the SRAM~\cite{roy2020memory}. 
Moreover, resistive memories such as Resistive RAM (ReRAM), magnetic RAM (MRAM), and phase-change memory (PCM), show other important advantages: non-volatility (NV), low power envelope, and multi-level storage~\cite{ 8310400}. IMC based on NV memories suffers from similar precision issues as SRAM-based IMCs, compounded by additional challenges coming from memristive devices, such as write variability and conductance variations (temporal and temperature-induced)~\cite{sebastian2019computational}.

From a system-level perspective, resistive memories serve not only as IMC primitives, but also as non-volatile storage blocks for DNN weights. This avoids moving weights across the system memory hierarchy, which is instead necessary for SRAM-based IMC. Contrarily to SRAM, re-programming the memristive cross-bars with new data during the network model execution is not affordable, due to the high latency and power consumption associated with re-writes of non-volatile memory cells, as well as their limited endurance (for ReRAM and PCM).
This forces rethinking architectures as memory-centric, with additional digital logic around to perform ancillary operations, as is the focus of this work.

Considering ReRAM-based IMC, Chen \textit{et al.}~\cite{8310400} demonstrate significant computing parallelism, performing 8k MAC operations simultaneously. Other works show ReRAM-based IMC arrays as dense as 2Mb~\cite{9063078} or 4Mb~\cite{9365769}, with peak energy efficiencies in the range of 120-200 TOPS/W within a power envelope of few milliwatts, suitable for tiny edge AI devices. Moreover, the IMC array in~\cite{9063078}, integrated within a PCB hosting also an FPGA, runs a ResNet-20 trained on the CIFAR-10 dataset with 90\% top-1 accuracy.

For the MRAM technology, Doevenspeck \textit{et al.}~\cite{doevenspeck2020sot} present a SOT-MRAM-based IMC macro and demonstrate for the first time that resistive MRAM devices can be used for DNN applications. They claim software-like accuracy on a network targeted to the MNIST dataset.

PCM-based IMC arrays have been applied in mixed-precision in-memory iterative computing, combining a computational memory unit to perform the bulk of a computational task, with a von Neumann machine, which implements a backward method to iteratively improve the accuracy of the solution. This approach has been demonstrated to solve linear equations~\cite{le2018mixed} and in DNN inference and even training tasks~\cite{sebastian2019computational}, showing limited error in the computation and much higher efficiency compared to traditional approaches~\cite{sebastian2020memory}.

Khaddam-Aljameh \textit{et al.}~\cite{hermesCor} recently presented a state-of-the-art 256$\times$256 PCM-based IMC core targeting DNN inference, fabricated in 14nm, showing energy efficiency of 10.5 TOPS/W and performance density of 1.59 TOPS/mm$^2$ on inference tasks of multi-layer perceptrons and ResNet-9 models trained on MNIST and CIFAR-10 datasets, with comparable accuracies as software baseline. In this work, we adopt the PCM-based IMC presented in~\cite{hermesCor}.

\subsubsection{Architectures and Systems}
As discussed, there are several challenges related to technology that affect both charge-based and resistive IMC circuits currently under scrutiny from researchers. However, provided that these issues can be solved, another essential challenge is the integration of in-memory computational primitives into heterogeneous systems.
In this work, we focus in particular on this aspect.

IMC cores primarily target matrix-vector multiplications (MVMs) or other similar vector operations, showing incredible throughput and efficiency. Although MVM operations are predominant in modern DNNs, they still represent only a subset of the DNN computation~\cite{InMComp}, which also includes residual connections, pooling layers, non-linear activation functions, softmax, etc.. Increasing the throughput of MVMs with IMC moves the performance bottleneck to all the other layers, which can not be easily mapped on IMC arrays. From a broader application perspective, an edge-computing system might incur workloads characteristically different than MVMs, such as data management and control tasks that are performed together with neural tasks~\cite{pulp_frontnet}. It is necessary, for a complete architecture, to address this computation in a programmable way. This reasoning strongly motivates the integration of the IMC with other specialized accelerators and software programmable cores in heterogeneous architectures~\cite{InMComp}.

To the best of our knowledge, not many works specifically focused on the integration of IMC arrays in heterogeneous analog/digital systems have been presented in literature so far. Dazzi~\textit{et al.}~\cite{ dazzi2021efficient} propose more advanced IMC multi-core approaches with very carefully staged core-to-core dataflow, but the focus is mostly on convolutions and there are no provisions for heterogeneous computing nor for computations that do not map efficiently on the AIMC arrays. Houshmand~\textit{et al.}~\cite{houshmand2020opportunities} explore co-optimization strategies of IMC array size, memory hierarchy and data-flows to avoid efficiency degradation when the IMC core is integrated into a processing infrastructure including also memory buffers and small control units, but they do not investigate complex scenarios like heterogeneous systems.

Zhou~\textit{et al.}~\cite{zhou2021analognets} propose a PCM-based IMC array modeled in 14nm technology, complemented with additional digital logic that performs activation and pooling operations. A small SRAM memory acts then as a layer-to-layer intermediate buffer, followed by a hardware block that handles \textsc{im2col} transformations. The proposed solution shows a peak 112 TOPS/W on MVMs and has been demonstrated on the execution of a custom DNN model, with  95.6\% of accuracy, at a performance of 7.7 inf/s with 8.22 $\mu J / inf$. However, this type of architecture is not flexible enough to support heterogeneous workloads, since it does not feature programmable cores.

Jia~\textit{et al.}~\cite{new_verma} propose a 4$\times$4 array of cores consisting of charge-based IMC cross-bars extended with a programmable near-memory-computing (NMC) digital accelerator that performs single-instruction-multiple-data (SIMD) computing, shifting, pooling, and activation functions. On 8-bit MVMs, the prototype in 16nm shows 3 TOPS of peak performance with 30 TOPS/W of efficiency. Coupled with off-chip FPGA and MCU that handle communication with a host PC and control flows, the prototype has been demonstrated on a ResNet-50 model with 4-bit weights and activations, achieving a peak performance of 3.4 TOPS.

The silicon prototype presented in~\cite{jia2020programmable} integrates a charge-domain compute-in-memory unit supporting 1to8-bit$\times$1to8-bit matrix-vector multiplications, into a tiny RISC-V CPU enriched with a direct memory access controller (DMA) and a set of peripherals. It shows a peak efficiency of 400 TOPS/W on the end-to-end inference of a binarized ten-layers network trained on CIFAR-10. However, also in this case the architecture can not afford complex heterogeneous computation: the core delivers only a few million operations per second and it can only be used for control tasks such as programming DMA transfers, not being capable of performing compute-intensive functions with sufficient performance level.

%
The limits of the above-mentioned systems are mainly two: the integration of the IMC accelerator (IMA) is loosely-coupled, with the IMA connected with other cores through a low bandwidth, high latency system bus; the presented heterogeneous systems are demonstrated either on neural networks model of few layers (trained on datasets such as CIFAR-10 or MNIST) or on custom NN models built ad-hoc to fit the requirements of the architecture. Neither approach is representative of modern DNN models widely used in classification and detection tasks at the edge of the IoT. 

In this manuscript, we extend the work we presented in~\cite{ottavi2021end} by coupling the IMA with a novel design of a digital accelerator to improve the efficiency of depth-wise kernels, integrated into the heterogeneous system and we fully implement the cluster in the GlobalFoundries 22nm FDX technology. Furthermore, we scale up the architecture and explore the resources necessary to enable the end-to-end inference of a full MobileNetV2 network, a much more realistic benchmark for the class of networks that an ultra-low-power IoT end-node could target.

%% file: text/03_background_v2.tex
\section{Background}
\begin{figure*}[t]
    \centering
    \includegraphics[width=0.85\linewidth]{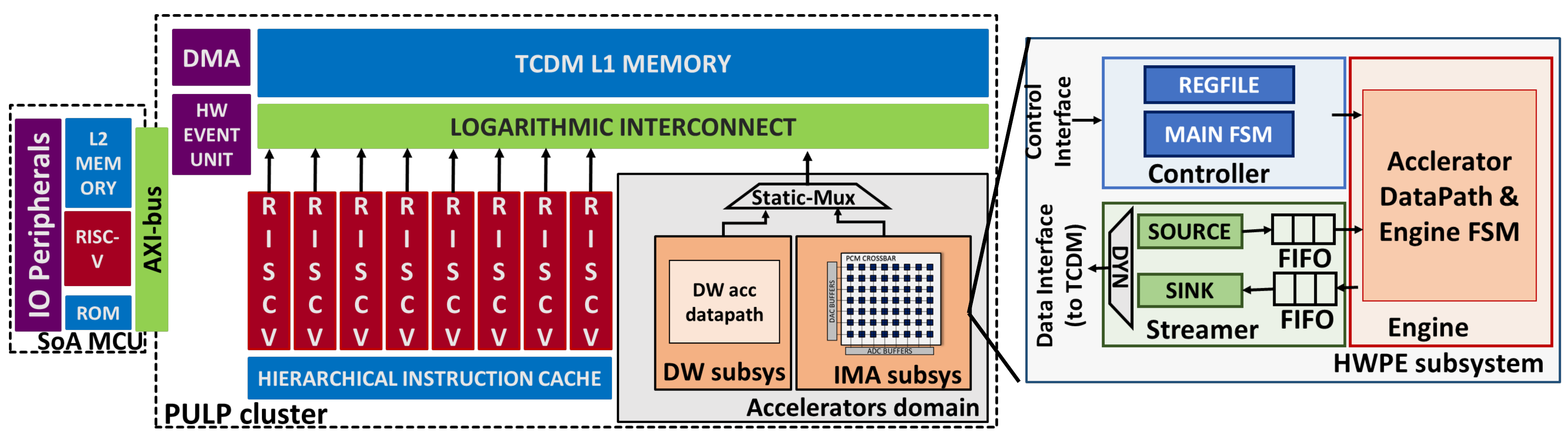}
    \caption{Overview of the PULP cluster architecture, integrating the IMA and the digital depth-wise accelerator. Each accelerator is enclosed into a Hardware Processing Engine (HWPE) subsystem, depicted on the right.}
    \label{fig:cluster_archi}
    \vspace{-2mm}
\end{figure*}
\label{sec:Background}

\subsection{PCM-based In-Memory Accelerator}
\label{sec:in-mem_comp}
In this paper, we use the SoA In-Memory Computing array presented in~\cite{hermesCor} which is based on a Phase-Change Memory (PCM) cross-bar. In this architecture, the memory devices are resistors with programmable conductance placed at the crosspoints of a 2D array with one terminal connected to horizontal wires called \textit{word-lines} and the other terminal connected to vertical wires called \textit{bit-lines}, enabling the execution of several computational primitives concurrently.

To perform the product of a matrix $\mathbf{A}$ by a vector $\mathbf{x}$, the PCM devices are programmed with conductance values proportional to the values $\mathbf{A}_{ij}$ of $\mathbf{A}$, with a precision of 4-bit (signed). Then the word-lines are driven with voltage pulses, whose duration are proportional to $\mathbf{x}_j$, using a set of digital-to-analog converters (DACs) with 8 bits of precision (signed). By Ohm's law, each PCM device contributes a current proportional to $\mathbf{A}_{ij}\cdot\mathbf{x}_j$ on the $i$-th bit-line, resulting in a total integrated current proportional to the dot product $\mathbf{y}_i = \sum_j \mathbf{A}_{ij}\cdot\mathbf{x}_j$. At the end of each bit-line, there is an analog-to-digital converter (ADC) used to sample the bit-line current and convert it into an 8-bit digital value (signed).

For DNN inference, the $\mathbf{A}$ matrix can be used to store the weights of the linear part of a Fully Connected, Convolutional, or Depthwise Convolutional layer. Note that typically 2 PCM devices are used to denote a signed weight~\cite{Y2020joshiNatComm}. In conventional digital architectures, the dot product of 4-bit weights and 8-bit input activations requires a high-precision intermediate representation (often, 32 bits) that is subject to scaling, clipping, and quantization to produce a vector of 8-bit output activations~\cite{burrello2020dory}. In the IMC cross-bar, instead, the intermediate representation is an analog current, while scaling, clipping, and quantization are performed directly by the bit-line ADCs by setting appropriate current limits. 

\subsection{The PULP Cluster}

Highly parallelizable workloads such as DNNs are a good fit for high core count heterogeneous systems that can integrate specialized accelerators.
The PULP cluster~\cite{rossi2021vega} we assume as a reference, depicted in Fig.~\ref{fig:cluster_archi}, incorporates 8 RISC-V cores, each featuring a 4-stage in-order single-issue pipeline and implementing the RISC-V RV32IMCXpulpV2 Instruction Set Architecture (ISA).  XpulpV2 is a custom extension to the RISC-V ISA~\cite{gautschi2017near} meant to accelerate arithmetic intensive kernels.

The cores of the baseline cluster communicate through a shared and word interleaved memory called Tightly Coupled Data Memory (TCDM), referred to as L1 memory. The size of the memory is parametrizable and, in this context, is 512 kB, divided on 32 banks. The cores access the memory through a low latency logarithmic interconnect (LIC), that serves the memory accesses in one cycle. 
The cluster workload can be offloaded to accelerators, integrated into the cluster through a standardized interface~\cite{conti2015ultra}, discussed in Sec.~\ref{sec:HWPE}. 

The cluster communicates with a micro-controller system that handles input/output peripherals, through an AXI interface. Moreover, it is served with a DMA controller dedicated to the data transfers between the TCDM and the second level of memory, hosted by the micro-controller system, which also contains the program instructions for the cluster cores. 
Each core fetches the instructions from a hierarchical instruction cache organized on two levels (the first private to each core, the second shared) to optimize the hit rate. 
The cluster is also supported by a Hardware Synchronization Unit that manages synchronization and thread dispatching, enabling low-overhead and fine-grained parallelism, thus high energy efficiency: each core or accelerator waiting for a barrier, or more in general for a custom event, is brought into a fully clock gated state.

%% file: text/04_body_v3.tex
\section{Heterogeneous System}
\label{sec:BODY}
In this section, we present the analog in-memory accelerator (IMA), the depth-wise digital accelerator, and their integration into the PULP cluster through a standardized interface called Hardware Processing Engine.

\subsection{Hardware Processing Engines}
\label{sec:HWPE}
To improve the performance and the energy efficiency of the accelerators in data movement operations, and to ease their integration into the cluster, each of the two accelerators presented here is incorporated as a Hardware Processing Engine (HWPE) using a standardized interface\footnote{https://hwpe-doc.readthedocs.io/en/latest/}. HWPEs expose a control and a data interface towards the rest of the cluster. The control interface allows the cluster's cores to access the registers of the targeted accelerator for configuration. The data interface is connected to the TCDM memory through multiple master ports on the logarithmic interconnect, similarly to what happens with the cores of the cluster. The width of this bus is a design-time parameter and can be chosen depending on the required bandwidth of the accelerator. 

Fig.\ref{fig:cluster_archi} shows the heterogeneous cluster with two distinct HWPE interfaces encapsulating the IMA (namely \textit{IMA subsystem}) and the depth-wise accelerator (namely \textit{DW subsystem}). To avoid a large increase in the area of the logarithmic interconnect and in the latency of its arbitration scheme, the data interface of the \textit{IMA subsystem} and the \textit{DW subsystem} are statically multiplexed towards the TCDM, sharing the same physical ports on the interconnect. The two accelerators are used in a time-interleaved fashion, allowing one accelerator to full access the TCDM at a time. This choice does not cause any performance degradation, since in our DNN computing model the depth-wise accelerator and the IMA can not be active concurrently. However, they can be programmed independently and in parallel by the cores of the clusters. Each accelerator has its own programming bus and the configuration registers are mapped in different regions of the cluster memory map. To ease the programming phase of the accelerators, we expose to the programmer a set of hardware-abstraction-layer (HAL) functions that can be inferred directly into the C code through their explicit invocations.
To reduce the power consumption of the cluster on jobs deployed to HWPEs, the latter expose an end-of-computation signal towards the cluster.
After programming the HWPE and triggering its execution, the cluster cores can enter a low-power clock-gated sleep mode. Once the HWPE notifies an end of computation signal, the core can be woken up by the cluster Event Unit.  

From the inside, HWPEs consist of three main blocks: the \textit{Controller}, the \textit{Engine}, and the \textit{Streamer}.
The \textit{Controller} contains a memory-mapped latch-based register file used to store the configuration of the execution of the accelerator, and the main \textit{Finite-State-Machine} (FSM) of the HWPE system, that coordinates the other blocks. The \textit{Controller} can be targeted by the cores of the cluster in a memory-mapped fashion via the control interface introduced above. The semantic and the number of registers as well as the FSM are customized to accommodate the requirements of the enclosed accelerator. 
The \textit{Engine} contains the data path of the accelerator and the specific FSM that coordinates the execution flow. It is, therefore, highly dependent on the specific accelerator design.

The \textit{Streamer} contains the blocks necessary to move inputs and results in and out of the accelerator through its master port of the data interface and transform the memory accesses into coherent streams to feed the accelerator \textit{Engine}. The streams are organized in two separated modules, namely \textit{source} for incoming streams and \textit{sink} for the outgoing ones. Both \textit{source} and \textit{sink} include address generators capable to generate three-dimensional access patterns in TCDM with configurable strides. They also include a re-aligner module to form word-aligned streams from non-word-aligned memory accesses, without constraining the memory system outside the HWPE to support misaligned accesses. The memory accesses generated by the two streams are dynamically multiplexed towards the data interface. Such a choice avoids the duplication of the data interface ports while not causing any performance overhead; eventual contentions are efficiently solved by an arbiter featuring a round-robin arbitration policy. Intermediate FIFOs in both directions are used to decouple the streams from memory contentions stalls and reduce the pressure on timing closure of the tightly-coupled system.

\subsection{IMA Subsystem Architecture}
\label{sec:execution_model}
\label{sec:ima_description}

\begin{figure}[t]
\centerline{\includegraphics[width=0.85\linewidth]{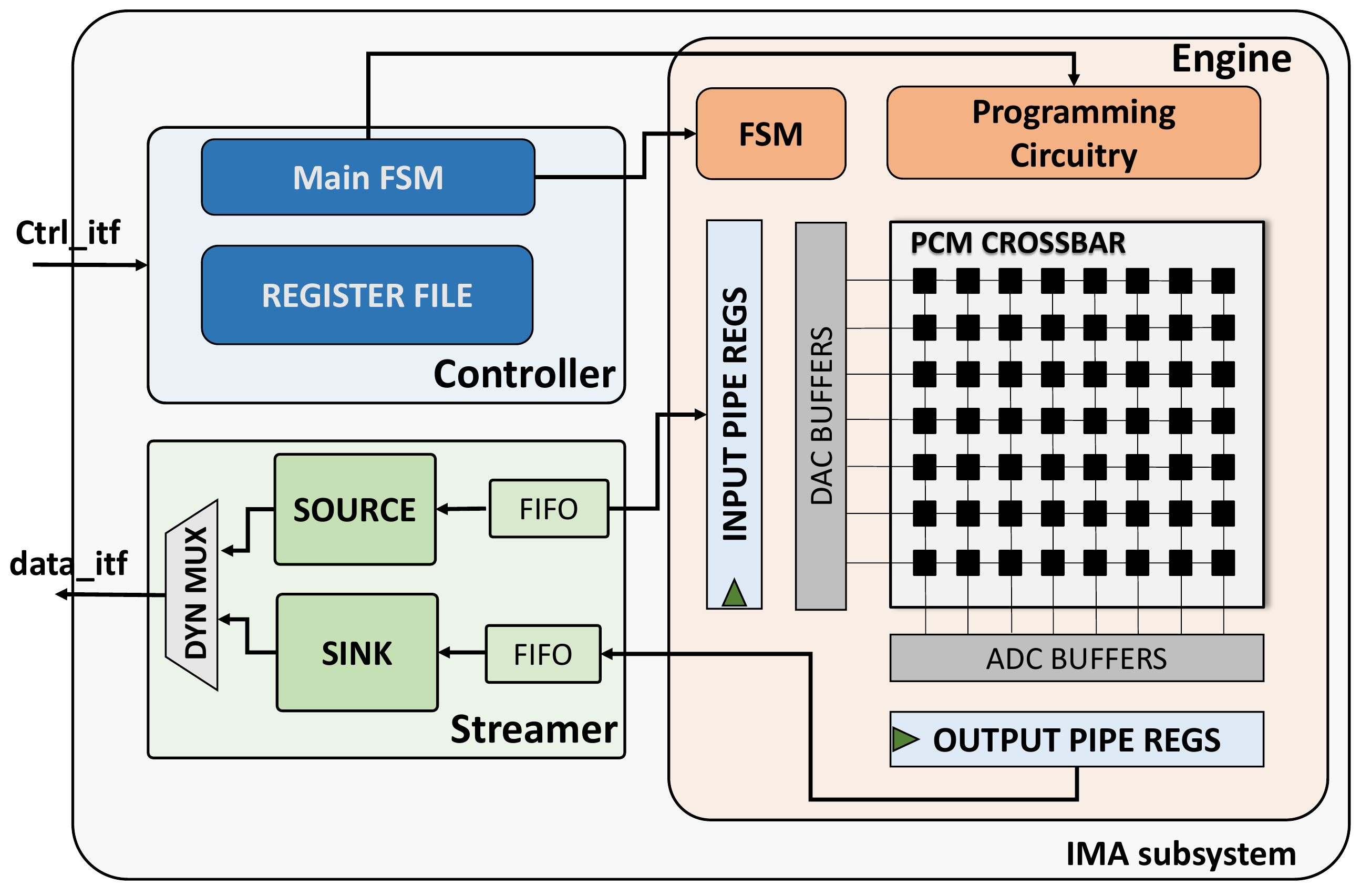}}
\caption{IMA enclosed into the HWPE interface. The \textit{data\_itf} width is designed to match the IO requirements of the IMA.}
\label{fig:ima_subsystem}
\vspace{-2mm}
\end{figure}

\begin{figure}[!t]
\centering
\includegraphics[width=0.9\linewidth]{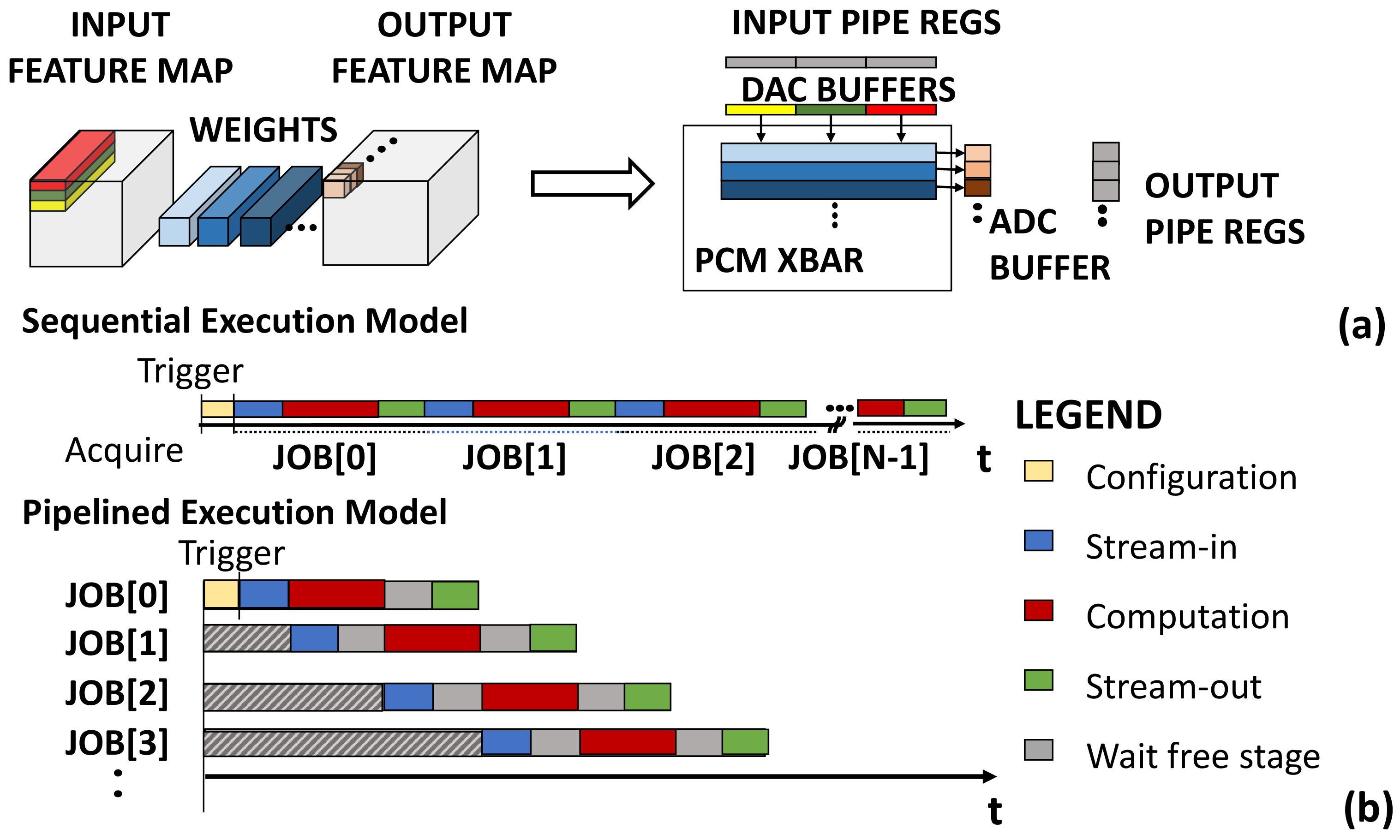}
\caption{(a) Mapping of standard convolutions on the PCM crossbar. (b) Timeline of the sequential and pipelined execution models.}
\label{fig:ima_map_plus_timeline}
\vspace{-2mm}
\end{figure}

Fig.~\ref{fig:ima_subsystem} shows the integration of the IMC cross-bar within the HWPE.
The width of the IMA data interface is sized to sustain the bandwidth requirements of the analog core, as shown in Sec.~\ref{sec:IMC_Accelerator_Performance}.
The \textit{Engine} contains both the digital and analog parts of the IMA data path. The digital part is composed of buffers for ADCs and DACs and of control circuitry; the analog core encloses the PCM devices (including PCM programming circuitry), and the ADCs and DACs themselves. 

The IMA works on input data stored in L1 with the HWC format, i.e., with consecutive data elements encoding pixels that are adjacent in the channel dimension. Fig.~\ref{fig:ima_map_plus_timeline} shows how a CNN layer is mapped on the IMA array.
For a standard convolution, the streamers can directly perform a virtual \textsc{im2col} transformation~\cite{garofalo2020pulp}, enabling to remap the computation to matrix-vector products of the form discussed in Sec.~\ref{sec:in-mem_comp}. As a consequence, the PCM array computes $C_{out}$ channels of one output feature map pixel from a complete input volume of $C_{in}\times K\times K$ pixels in a single operation (that we call \textit{job}), where $C_{in}$, $C_{out}$ indicate the number of input and output channels, and $K$ is the filter size.

The configuration sequence of the IMA starts when a core acquires a lock over the accelerator by reading a special \textsc{acquire} register through the control interface. After that, the core can interact with the IMA by programming the PCM devices with the weights of one or multiple layers; reading the conductance value of a PCM device; configuring a \textit{job} setting the address of input and output data in TCDM and the ADC configuration. When the configuration ends, the execution can be started by writing to a special \textsc{trigger} register. To minimize the IMA configuration and synchronization overhead, multiple jobs can be pipelined by setting the register file with the correct strides. In this way, a whole layer can be executed with only one configuration phase.

We propose two execution models for back-to-back job operations of the IMA: a simpler one, sequential, and a more optimized pipelined execution model. The relative timelines are shown in Fig.~\ref{fig:ima_map_plus_timeline}. The sequential model splits the execution of the single job into three phases operated sequentially. \textsc{stream-in}: fetch data from the TCDM that is then streamed to the engine's internal DACs buffers; \textsc{computation}: analog computation on the crossbar and writing of the ADCs buffers; \textsc{stream-out}: stream data from buffers back to the TCDM. 
In Sec.~\ref{sec:IMC_Accelerator_Performance}, we study how this model quickly becomes a bottleneck for the IMA's peak performance.

In the pipelined execution model, the three aforementioned phases of different jobs can overlap each other at the cost of additional hardware resources: we add two pipeline registers before and after the DACs and ADCs buffers and we extend the \textit{Engine} FSM with additional states to control the overlapping phases: during the computing phase of the $i-th$ job (if not the last one to compute), the \textit{engine} FSM sets the streamer to start a new memory transaction to fetch the inputs for the successive $(i+1)-th$ job. When such stream-in phase has finished, if there are the results of the previous job $(i-1)-th$ to stream-out, the \textit{engine} FSM can configure the stream, as shown in Fig.~\ref{fig:ima_map_plus_timeline}. If we consider only the digital logic of the accelerator around the IMA, the pipelined approach increases the area by about 40\%, due to the doubled number of input/output registers needed to enable the pipeline. However, this overhead reduces to 5\% if we consider the total area of the accelerator (digital logic and analog IMC cross-bar), compared to the sequential approach. 



\subsection{Dedicated Depth-Wise Digital Accelerator}
\label{sec:dw_accelerator_archi}

\begin{figure}[t]
    \centering
    \includegraphics[width=0.9\linewidth]{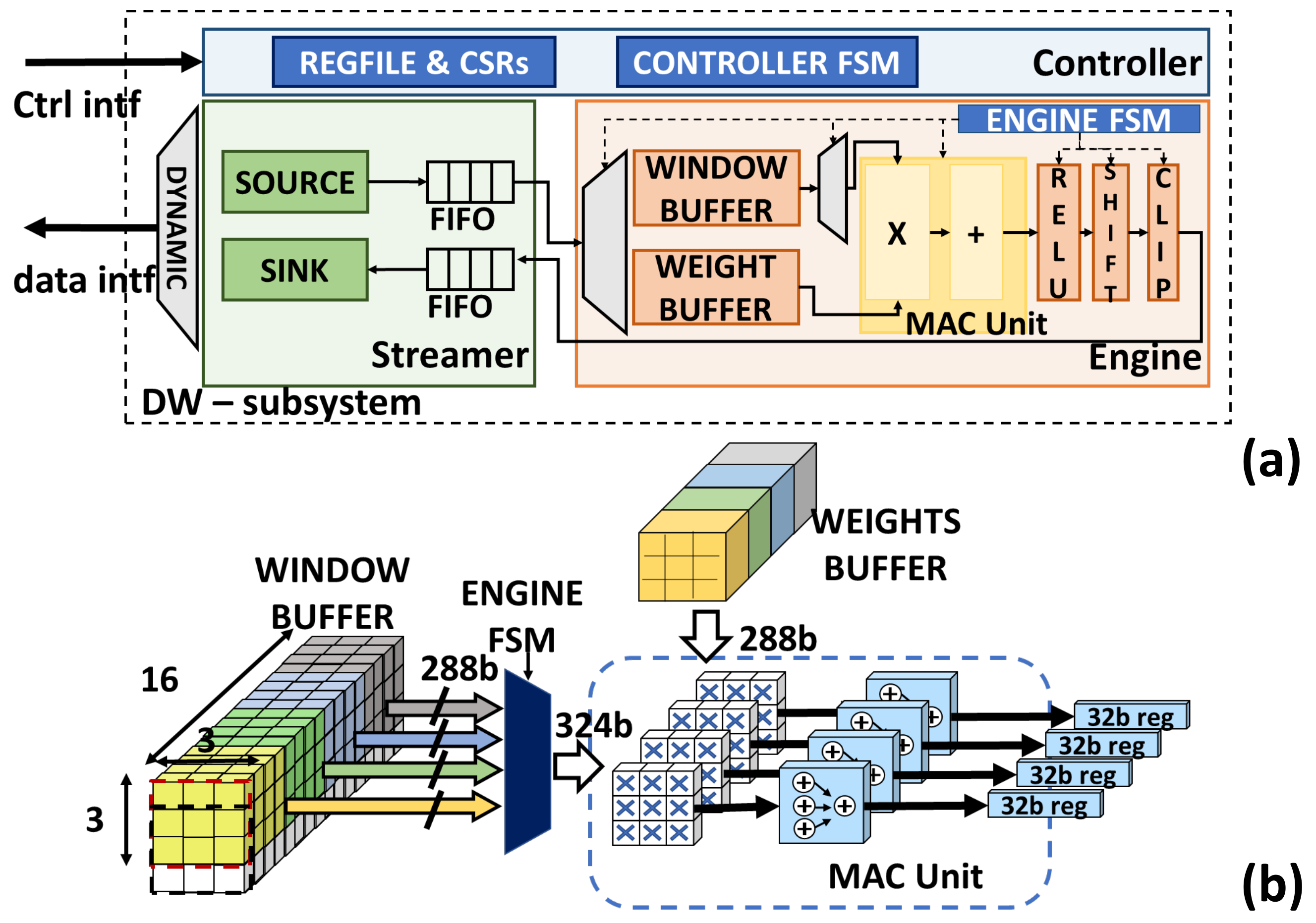}
    \caption{(a) Architecture overview of the Depth-wise digital accelerator, enclosed in the HWPE. (b) Execution flow of the depth-wise operation.}
    \label{fig:dw_archi}
\end{figure}
\begin{figure}[t]
    \centering
    \includegraphics[width=0.9\linewidth]{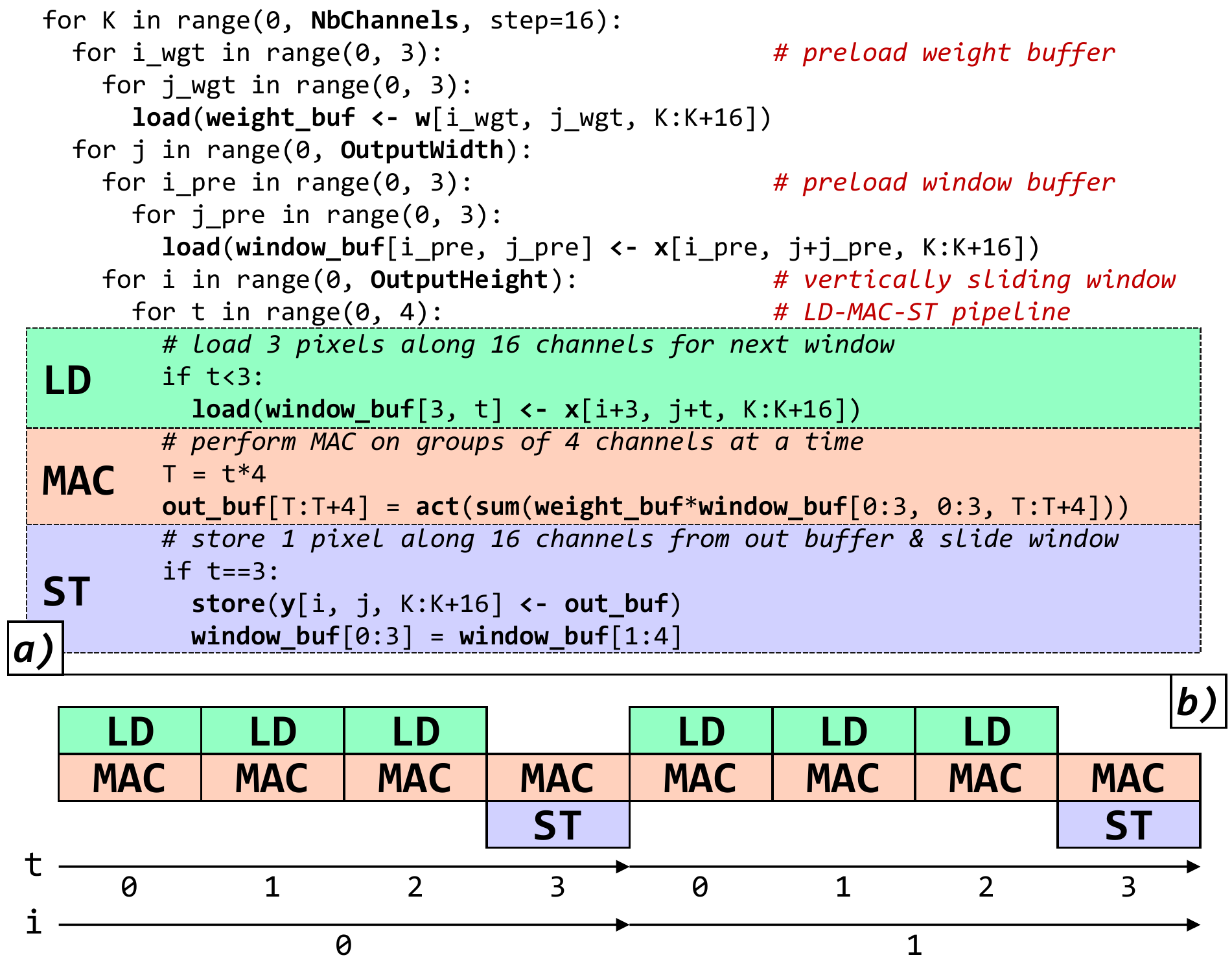} 
    \caption{(a) Pseudo-Python code describing the operation of the depth-wise accelerator datapath. (b) Detail of the \textit{LD - MAC - ST} pipeline.}
    \label{fig:DW_loops_pipeline}
    \vspace{-2mm}
\end{figure}

Depth-wise (DW) convolutions have been introduced in SoA DNNs such as MobileNetV2 to shrink the model size of the neural networks ( by 7 to 10$\times$) and their computational cost, with negligible accuracy drop~\cite{sandler2018mobilenetv2}. Due to their lower connectivity compared to standard convolutions (each output channel depends only on a single corresponding input channel), DW layers are generally inefficient to map on IMC arrays, as we show in Sec.~\ref{sec:IMC_Accelerator_Performance} on the \textit{Bottleneck} use-case. Moreover, a pure software execution of such kernels easily becomes a performance bottleneck for computation~\cite{ottavi2021end}. To speed up the execution of the depth-wise layers, we therefore designed a specialized digital accelerator and integrated it into the heterogeneous cluster.

The accelerator we propose in this work is capable of processing depth-wise kernels on 8-bit signed input tensors and weights, accumulating the results in intermediate 32-bit registers and performing non-linear activation functions such as ReLU plus a set of ancillary functions (i.e. shifting and clipping) to bring back the final result into the 8-bit precision.
Fig.~\ref{fig:dw_archi} shows the general architecture of the proposed accelerator, enclosed in the dedicated HWPE interface.
%
%
%

The depth-wise accelerator employs a weight-stationary data flow and targets 3x3 depth-wise layers -- the ones most commonly encountered in DNNs.
The weights from 16 different filters, assumed to be 8-bit signed elements, are loaded from the TCDM memory at the beginning of the computation, sign-extended, and stored into a \textit{weight buffer} that features 3x3x16 registers. 
The weights reside in the buffer until they have been used over the full input image.
Input tensors are scanned by the accelerator using a vertically sliding window on the spatial dimensions, considering in each iteration 16 channels data stored in HWC layout (i.e., the same layout used by the IMA).
The vertically sliding window is implemented utilizing a \textit{window buffer} of 4x3x16 8-bit registers: 3 rows to host the current window, plus 1 row of inputs being loaded concurrently with the current window computation.
Other than the two buffers, the data path of the accelerator consists of a network of Multiply-and-Accumulate (MAC), and ancillary blocks to compute ReLu, shifting and clipping operations, as shown in Fig.~\ref{fig:dw_archi}.
The MAC unit consists of 36 multipliers and a reduction tree that operate on a 3$\times$3$\times$4 block of the window buffer, passed through the ReLu and shifting\&clipping blocks, and stored in an \textit{output buffer}.
\begin{figure*}[t]
    \centering
    \includegraphics[width=0.85\linewidth]{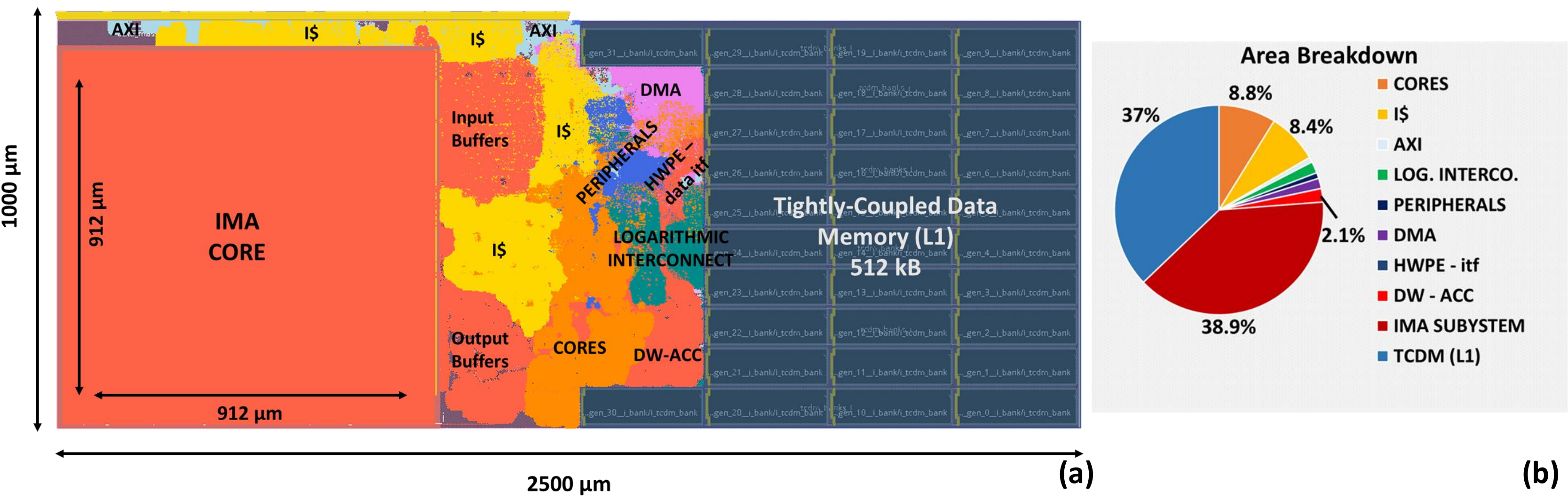}
    \caption{(a) Placed and Routed design of the heterogeneous cluster. (b) Area breakdown of the system.}
    \label{fig:floorplan}
    \vspace{-2mm}
\end{figure*}
Fig.~\ref{fig:DW_loops_pipeline}a shows the details of the depth-wise accelerator datapath operation in the form of Python-like pseudocode.
For a given block of 16 channels, the operation starts by preloading weights.
At the start of each output column, the window buffer is loaded with the content of the first 3x3 window; then, the operation of the datapath is organized in three pipelined stages, active over an inner loop of 4 cycles as shown in Fig.~\ref{fig:DW_loops_pipeline}a and Fig.~\ref{fig:DW_loops_pipeline}b.
In the first three cycles of the inner loop, the \textit{LD} stage is active: one input pixel across 16 channels is loaded to fill the fourth row of the window buffer.
The \textit{MAC} stage is active in all cycles of the inner loop, working on 4 channels at a time.
Finally, the \textit{ST} stage is active only in the fourth cycle: during this stage, the content of the output buffer produced in the previous three cycles and the current one is streamed out of the datapath, and the window buffer slides one pixel down.
In this way, during the main body of the computation, the accelerator fully exploits the available memory bandwidth of 16 Bytes per cycle and the HWC layout of data, which is advantageous because it is the same layout used by the IMA.
Overall, the execution of a depth-wise layer on the dedicated accelerator improves by 26$\times$ over a pure software implementation, achieving an average performance of 29.7 MAC/cycle.

%% file: text/05_results_and_discussion_v2.tex
\section{Experimental Results}
\label{sec:Experimental_Results}
This section evaluates the proposed heterogeneous cluster. From a physical viewpoint, we analyze area, power and timing costs of the  system. From a performance and energy efficiency viewpoint, we report the results of benchmarking hetereogeneous DNN layers, such as the \textit{Bottleneck}.

\label{sec:Physical_Implementation}


\subsection{Physical Implementation}
To characterize the system in terms of area, power, and performance, we implement the cluster using the GlobalFoundries 22nm FDX technology node. We synthesize the heterogeneous cluster with Synopsys Design Compiler-2019.12 and we perform a full place\&route flow using Cadence Innovus 20.12, targeting the worst-case corner (SS, 0.72V, -40°/125°). The floorplan of the system is reported in Fig.~\ref{fig:floorplan}. The analog IMC accelerator models, validated on silicon and modeled through silicon characterization of 14 nm prototypes, are fed into technology libraries (\textit{.lef}, \textit{.db}, and \textit{.lib}) integrated into the front-end and back-end flows of the system. The area, the timing, and the power consumption of the IMC accelerator are extrapolated from the on-chip measurements reported in~\cite{hermesCor}, properly scaled to the 22nm technology node. The power scaling is done according to the classical scaling theory under constant frequency, scaling power by $a\cdot b ^2$, where $a$ denotes the dimensional scaling and $b$ is the voltage scaling factor. The area scaling follows the dimensional scaling. We assume that the IMA latency will remain constant. The total area of the heterogeneous cluster is 2.5 $mm^2$, partitioned among the several hardware blocks as shown in Fig.~\ref{fig:floorplan}(b). As expected, the IMA sub-system and the 512 kB of TCDM memory occupy the major part of the total area ($\sim$1/3 IMA, $\sim$1/3 TCDM, 1/3 the rest of the cluster), while the depth-wise accelerator has a negligible impact (2.1 \%). The maximum operating frequency achievable by the final design is 500 MHz.

To perform power measurements we run parasitics-annotated gate-level netlist simulations of the digital part of the system in the typical corner (TT, 25C) at the operating voltage of 0.8V, executing DNN layers introduced above. The VCD simulation traces are analyzed with the Synopsys PrimeTime tool and the extracted power is integrated with the power extrapolated for the IMC accelerator~\cite{hermesCor}.
%
Hence, the results presented in this section and in the following ones include the overheads (i.e. timing, area, power) caused by the clock tree implementation, accurate parasitic models extraction, cell sizing for setup fixing, and delay buffers for hold fixing.
We emphasize that neglecting these factors would cause significant underestimations in the clock tree dynamic power.

\begin{figure*}[!t]
    \centering
    \includegraphics[width=0.9\linewidth]{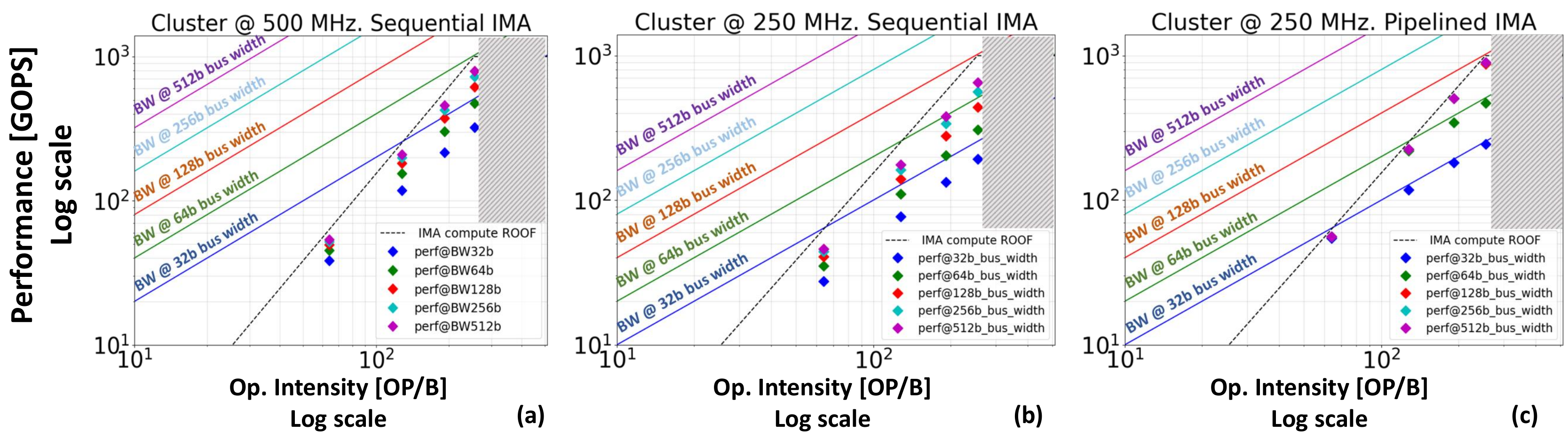}
    \caption{Roofline plot of the IMA heterogeneous system. The compute roof of the IMA is a diagonal line, quadratically dependent on operation intensity, not on cluster frequency. The intersection of a bandwidth line with the compute roof defines a region where performance points can lay for that configuration. In (a) and (b) cluster runs at 500MHz and 250MHz, with sequential execution of IMA. In (c) it runs at 250 MHz with a pipelined execution model for the IMA.}
    \label{fig:roofline}
    \vspace{-2mm}
\end{figure*}
\subsection{IMC Accelerator Performance}
\label{sec:IMC_Accelerator_Performance}
First, we analyze the peak performance achievable by the IMA. An important point to stress is that, in contrast with digital accelerators, the maximum performance of the IMC array is only related to its MVM operation latency and its size (256$\times$256 in the context of this work). The peak throughput is 1.008 TOPS, given by the maximum number of operations (256 $\times$ 256 $\times$ 2 OPs) that can be executed in its latency of 130ns~\cite{hermesCor}. The real throughput achievable is typically scaled by the utilization factor of the array: only if we map a 256 output-channel / 256 input-channel point-wise layer we can achieve the maximum utilization rate and, thus, throughput.
Another factor that limits the performance of the IMA sub-system is the memory bandwidth that the heterogeneous cluster can sustain to feed the IMA with new input data and to store the IMA results into the TCDM memory. If the computation is too fast compared to the stream-in and stream-out time, we lose performance because we are limited by the bandwidth of the system. 

The PULP cluster potentially offers high memory bandwidth towards the TCDM thanks to the tightly-coupled interconnect scheme, at the cost of increased interconnect area (linearly scaling with the bit-width of the system bus), power and timing. 
To find the width of the system bus able to sustain the IO requirements of the IMA at the lowest area overhead, we benchmark synthetic point-wise layers with different utilization rates of the IMC array (from 5\% to 100\%), varying the width of the IMA sub-system bus from 32- to 512-bit.

In Fig.~\ref{fig:roofline} we report the outcomes of our exploration as a roof-line plot~\cite{williams2009roofline}. The computing latency of the IMA does not depend on the cluster frequency, leading to two considerations: first, the compute roof of the IMA is a diagonal line proportional to the operation intensity (in other words, to the utilization factor of the IMA cross-bar) and not a line parallel to the x-axis, as is typically the case for digital systems; second, since the IMA computing latency is fixed, its memory bandwidth requirement change as we reduce or increase the cluster clock frequency. We investigate two operating frequencies, the maximum achievable one by the system when operating at high-voltage (500 MHz at 0.8V) and the maximum one achievable at low-voltage (250 MHz at 0.65V), and we compare the sequential and the pipelined execution models of the IMA. 

In Fig.~\ref{fig:roofline}(a) the cluster is running at 500 MHz and we adopt the sequential execution model for the IMA. We observe that only with a 32-bit wide bus we are memory bound and a 64-bit wide data interface of the IMA subsystem is sufficient to fulfill the computing and IO requirements of the IMA. However, analyzing the performance at any of the system bus configurations above 32-bit we notice a gap between the compute roof and the real throughput, suggesting that we are under-utilizing the bandwidth of the system. The reason is that in the sequential model, as discussed in Sec.~\ref{sec:ima_description}, 8\% to 40\% (depending on the size of the layer considered) of the total execution cycles are spent in stream-in and stream-out phases. In the rest of the execution, when the IMA is in the computing phase, the system bus is not used.

Analyzing the scenario where the cluster runs at 250 MHz, Fig.\ref{fig:roofline}(b), we observe that higher bus-width (i.e. 128-bit) is necessary to preserve the peak performance of the IMA. However, also in this case the sequential execution model is quite far from reaching the computing roof of the IMA.

Despite the unavoidable area overhead compared to the sequential execution model (which, however, is limited to 5\% if we consider the whole IMA sub-system, including the analog macro), Fig.\ref{fig:roofline}(c) shows that the pipelined solution guarantees full utilization of the bandwidth. At the system level, the optimal configuration is with a 128-bit wide system bus: using a narrower system bus, throughput is memory-bound, while using a wider one, performance does not improve, as computation is in a  compute-bound region. In the optimal system configuration, the IMA can achieve a peak of 958 GOPS at 250 MHz, only 10\% less than the theoretical peak performance at the compute roof, due to the programming overhead to configure the accelerator and start the execution.

\subsection{Case Study: The \textit{Bottleneck} Layer}
\label{sec:The_Bottleneck_Layer}
\label{sec:ima_depthwise}
\label{sec:bottlenecs_res}

\begin{figure}[t!]
\centerline{\includegraphics[width=0.9\linewidth]{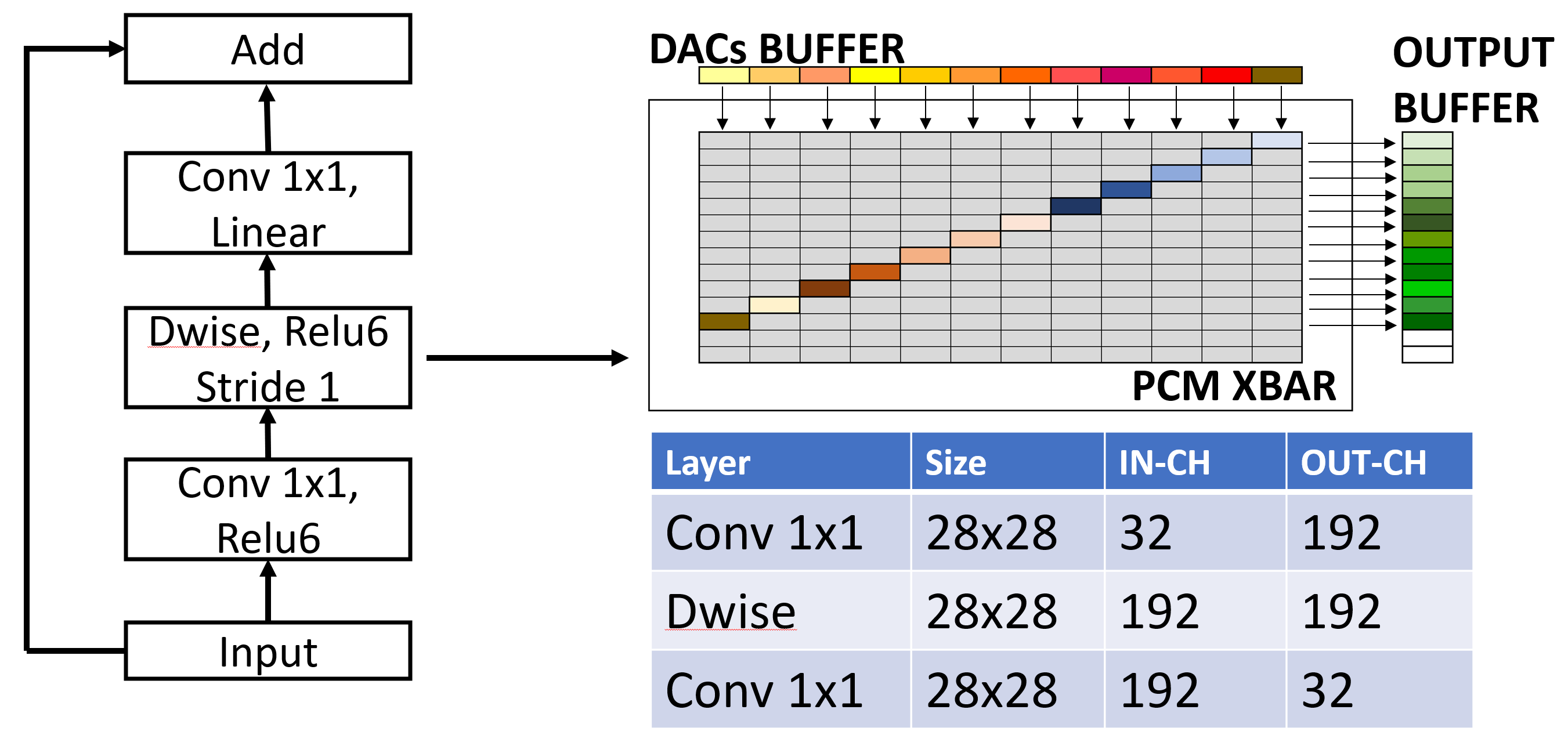}}
\caption{Parameters of the \textit{Bottleneck} layer and mapping structure of the depth-wise on the PCM crossbar. Gray rectangles are padding required for computing more than 1 channel per job.}
\label{fig:bottleneck}
\vspace{-2mm}
\end{figure}
\begin{figure*}[t]
\centerline{\includegraphics[width=0.85\textwidth]{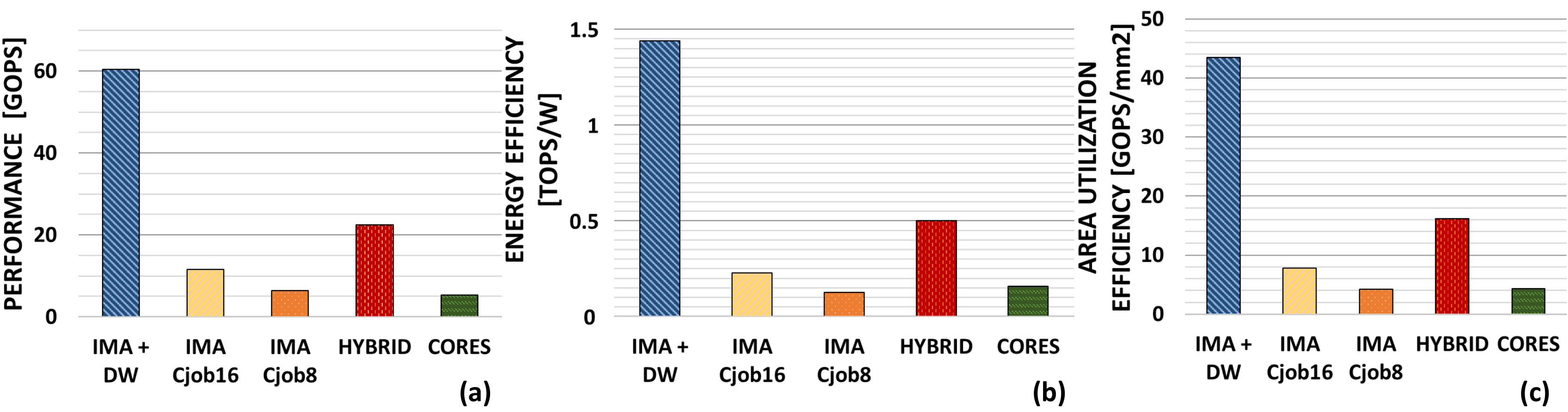}}
\caption{(a) Performance (in GOPS), (b) Energy Efficiency (in TOPS/W), and (c) Area Utilization Efficiency (in GOPS/$mm^2$) of the \textit{Bottleneck} layer running on the cluster at 500 MHz with 128-bit wide system-bus. The area efficiency is related to the effective area of the PCM arrays utilized to implement the \textit{Bottleneck} (including padding necessary to map the depth-wise on the IMA).}
\label{fig:res_compressed}
\vspace{-2mm}
\end{figure*}
\begin{figure}[t]
\centering
\includegraphics[width=0.9\linewidth]{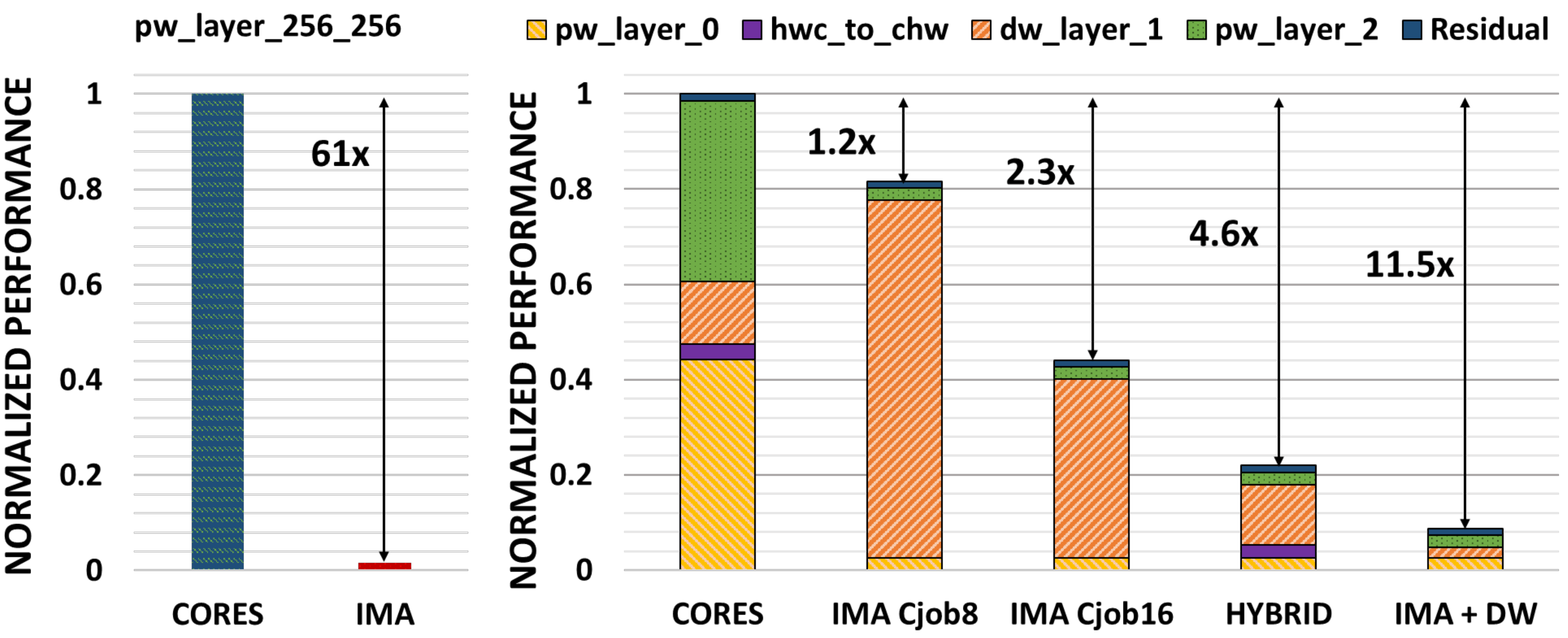}
\caption{Normalized Performance, compared to full software implementation (\textsc{cores}), of point-wise (left) and \textit{Bottleneck} (right) layers. For the right-side analysis, the impact of each layer on the execution of the whole Bottleneck is shown, considering the different computing mapping solutions enabled by the heterogeneous cluster.}
\vspace{-2mm}
\label{fig:totCycles}
\end{figure}

To highlight the advantages, the trade-off, and the challenges of IMC on realistic use cases for edge computing and to assess the benefits of the presented heterogeneous system, we benchmark the \textit{Bottleneck} layer of a MobileNetV2 DNN. We analyze three different computation mappings that are possible on the analog/digital system, compared to the baseline, which executes all the layers of the \textit{Bottleneck} on the software cores using optimized software libraries~\cite{garofalo2020pulp}.
The parameters of the selected \textit{Bottleneck} layer are reported in Fig. \ref{fig:bottleneck}: this configuration is chosen so that all the weights and activations fit the on-cluster TCDM (512 kB), without requiring any activation data tiling~\cite{burrello2020dory}, the in-depth study of which is beyond the scope of this work.

The first possible execution mapping is to offload all the layers of the \textit{Bottleneck} to the IMA accelerator, except  for the residual connection, which is always offloaded to the cluster's cores. To map the weights of the layers on the IMC cross-bar, we adopt the \textsc{im2col} approach~\cite{garofalo2020pulp}. Mapping point-wise layers is straightforward: each $1 \times 1 \times C_{in}$ filter is mapped across the height of the cross-bar (one column), more filters are mapped across the columns. If the layer parameters did not fit the size of the array, we would have to split the weights over multiple IMAs. We postpone the analysis of more complex scenarios, such as these, to Sec.~\ref{sec:discussion} and we focus on the baseline case of a fully fitting layer here.

Contrarily to the point-wise layer, the depth-wise one is very inefficient to map on the IMA cross-bar. In depth-wise convolutions, each output channel depends only on the corresponding input channel: to make them suitable for mapping on the cross-bar array, a $K \times K$ kernel with $C$ in/out channels must be expanded into a dense form, with all the weights out of a diagonal set to zero (padding), as shown in Fig.~\ref{fig:bottleneck}. Assuming a hypothetical IMC array large enough to fit all the weights and padding of the layer, out of $K^2 \times C^2$ crossbar locations programmed with weights or zeros, only $K^2\times C$ of them would concur to the kernel computation.

To reduce the useless occupancy of crossbar cells (i.e. programmed with zeros), a different approach is to split the computation of $C_{out}$ pixels, that normally would be computed in a single operation (what we call \textit{job}), over multiple jobs, each of which computes  $C_{job}<C_{out}$ pixels. As a trade-off, this leads to a smaller amount of operations per job, reducing the overall performance. The advantage of this method is that the total number of crossbar elements required to map the depth-wise kernel is $N_{xbar} = K^2 \times C \times C_{job}$, reducing the number of the overall programmed cells (with zeros and weights) by a factor of $C_{out} / C_{job}$ compared to the previous approach, at the cost of additional $N_{jobs}=C_{out}/ C_{job}$ jobs per output pixel to complete the execution of the kernel (note that in the previous case 1 job per $C_{out}$ pixels is possible only on ideal infinite sized cross-bar).

Mapping all the layers of the targeted \textit{Bottleneck} following the first approach is not feasible on the 256$\times$256 cross-bar array we use: we would require 23$\times$ more cross-bar locations than the real number of weights, running out of IMC resources. Hence, we analyze the costs of separating the depth-wise in multiple jobs, considering two parameters: $C_{job} = 8$ and $C_{job} = 16$, which translates to an increase of 25\% and 54\% in the number of devices, respectively. Empirically, we consider these configurations as a reasonable trade-off between performance and occupancy of the cross-bar. The two configurations are referred to as \textsc{ima $c_{job8}$} and \textsc{ima $c_{job16}$}, respectively.

The second mapping we analyze executes the point-wise layers on the IMA and the depth-wise kernels on the 8 RISC-V cores of the cluster. The software kernels for the depth-wise are derived from an optimized parallel software library tailored on PULP-based clusters~\cite{garofalo2020pulp}. Since such kernels require the input data to be in Channel-Height-Width (CHW) layout and since the output from the point-wise layer (from the IMA) is in Height-Width-Channel (HWC) layout, additional execution cycles are needed for on-the-fly data marshaling operations. The output is generated in the HWC format instead and can be forwarded to the IMA with no additional overhead. This configuration is referred to as \textsc{hybrid} and requires the storage of depth-wise weights in the TCDM, instead of in the IMA crossbar. This is a reasonable trade-off since the depth-wise weights usually account for no more than 10\% of the total of a depth-wise based neural network~\cite{sandler2018mobilenetv2} ($\sim4\%$ in the considered \textit{Bottleneck} layer). 

The third mapping solution, indicated as \textsc{ima+dw}, runs the point-wise layers on the IMA, the depth-wise layers on the dedicated digital accelerator, and the residual layer on the cores. The digital accelerator accepts input data and weights in HWC format and produces outputs in HWC format, requiring no additional data marshaling operations during the \textit{Bottleneck} layer execution. 

Benchmarking results are provided in Fig.~\ref{fig:res_compressed} for all the solutions discussed above, in terms of performance, energy efficiency, and area utilization efficiency. The width of the system bus is 128-bit and we adopt the pipelined execution model for the IMA; as demonstrated in Secion~\ref{sec:IMC_Accelerator_Performance}, this configuration maximizes performance. The cluster operates at 500 MHz at 0.8V, in typical operating conditions (TT, 0.8V, 25C). 

We can notice that despite a significant area utilization of the IMC array, the performance of \textsc{ima $c_{job16}$} and \textsc{ima $c_{job8}$} are only 2.27$\times$ and 1.23$\times$ higher than a pure software execution of the \textit{Bottleneck}. Efficiency is even worse: 1.23$\times$ lower energy efficiency and the same area efficiency of \textsc{ima $c_{job8}$}, and comparable energy and area efficiency of \textsc{ima $c_{job16}$} compared to the \textsc{cores} demonstrate that IMC arrays are not efficient to host sparse layers like the depth-wise. 
The \textsc{hybrid} solution instead achieves 4.6$\times$ better performance and 3.4$\times$ better energy efficiency than the \textsc{cores} configuration. Despite software-based execution of depth-wise layers, this solution surpasses the $c_{job16}$ configuration by 2$\times$ in terms of performance, by 2.3$\times$ in terms of energy efficiency, and by 2.1$\times$ in terms of area efficiency. The peak performance is achieved in the \textsc{ima+dw} configuration, improving by 2.6$\times$ and 11.5$\times$ over the \textsc{hybrid} and \textsc{cores} solutions, respectively. By offloading point-wise layers to the IMA and depth-wise layers to the dedicated digital accelerator, we fully exploit the potential of the two hardware blocks, while the cores handle their configuration, the workload dispatching, and ancillary aggregation operations, such as the residual connection. This synergistic approach, enabled by the fact that cores, IMA, and depth-wise accelerator all share the same memory at L1, stands out also as the most efficient one, achieving 2.7$\times$ and 9.2$\times$ improvements with respect to the \textsc{hybrid} and \textsc{cores} configurations, in terms of energy efficiency, and by 2.5$\times$ and 10.2$\times$ the same configurations in terms of area efficiency.

Fig.~\ref{fig:totCycles} shows the execution breakdown of the \textit{Bottleneck} layer. In a pure software execution scenario, the point-wise layers dominate the computation (\textsc{cores}). The IMA shows significant acceleration in such dense MVM-based operations (left-sided Fig.~\ref{fig:totCycles}), moving the performance bottleneck on other layers like the depth-wise. However, the IMA itself is not capable of mitigating this Amdahl's effect, since the execution of the depth-wise on the IMA is not efficient and dominates the total execution cycles (\textsc{ima $c_{job8}$} and \textsc{ima $c_{job16}$}). Execution of depth-wise convolutions on the cores (\textsc{hybrid}) improves execution time, but this block remains by far the slowest one. On the other hand, offloading the depth-wise layer to the digital accelerator (\textsc{ima+dw}) eliminates the performance bottleneck as the execution time of the depth-wise layer is comparable to the other components of the \textit{Bottleneck}, such as point-wise layers and residuals.

%% file: text/06_MobileNetV2_inference.tex
\section{End-to-End MobileNetV2 Inference}
\label{sec:discussion}

\begin{algorithm}
    \footnotesize
    \caption{Full-Network Tile\&Pack algorithm}
    \label{alg:tile_pack}
    \begin{algorithmic}[1] 
        \Function{Tile\&Pack}{$\mathbf{n}, \mathbf{h}, \mathbf{w}, S, n_\mathrm{ima}$} \Comment{$\mathbf{n}, \mathbf{h}, \mathbf{w}$ are name, height, width of all layers, $S$ is the size of each IMA (default 256), $n_\mathrm{ima}$ is the number of available IMAs}
            \State $\mathbf{Tiles} \gets [\,]$
            \ForAll{$n,(h,w) \in \mathbf{n}, (\mathbf{h}, \mathbf{w})$} \Comment{Create tiles}
                    \State $n_\mathrm{tile,w} \gets \lfloor w / S \rfloor$ \;;\; $w_\mathrm{rem} \gets w \bmod S$
                    \State $n_\mathrm{tile,h} \gets \lfloor h / S \rfloor$ \;;\; $h_\mathrm{rem} \gets h \bmod S$
                    \For{$i \in [0, n_\mathrm{tile,h}-1]$}
                        \For{$j \in [0, n_\mathrm{tile,w}-1]$}
                            \State $\mathbf{Tiles}[n+\mathrm{``\_tile}\,i\_j\mathrm{"}] \gets (S,S)$
                        \EndFor
                    \EndFor
                    \For{$j \in [0, n_\mathrm{tile,w}-1]$}
                        \State $\mathbf{Tiles}[n+\mathrm{``\_tile}\,n_\mathrm{tile,h}\_j\mathrm{"}] \gets (h_\mathrm{rem},S)$
                    \EndFor
                    \For{$i \in [0, n_\mathrm{tile,h}-1]$}
                        \State $\mathbf{Tiles}[n+\mathrm{``\_tile}\,i\_n_\mathrm{tile,w}\mathrm{"}] \gets (S, w_\mathrm{rem})$
                    \EndFor
                    \State $\mathbf{Tiles}[n+\mathrm{``\_tile}\,n_\mathrm{tile,h}\_n_\mathrm{tile,w}\mathrm{"}] \gets (h_\mathrm{rem},w_\mathrm{rem})$

            \EndFor\label{tilingend}
            \ForAll{$n,(h,w) \in \mathrm{Tiles}$} \Comment{Remove 0-sized tiles}
                \If{$h=0$ or $w=0$}
                    remove($\mathbf{Tiles}[n]$)
                \EndIf
            \EndFor; \quad
            \State $\mathbf{Bins} \gets \Call{BinBestFit}{\mathbf{Tiles}}$
            \State $\mathbf{IMA\_Mapping} \gets \Call{MaxRectsBSSF}{\mathbf{Bins}}$
            \State \Return $\mathbf{IMA\_Mapping}$
        \EndFunction
    \end{algorithmic}
\end{algorithm}

In this section, we study the scalability of the heterogeneous system (in terms of challenges and hardware resources) to enable end-to-end inference of the  MobileNetV2~\cite{sandler2018mobilenetv2}.
To build the model of the scaled-up architecture, we start from the physical measurements carried out in the previous sections, introducing the following considerations:
the PCM-based IMC cross-bar we use in this work does not support cell re-programming, during the execution of the Neural Network model, due to the high latency of the operation. An iterative flow is necessary to program each cell of the PCM cross-bar: first, pulses are sent to the cell, then the conductance is read-out and compared with the expected value. The outcome discrepancy is used to modulate the successive pulses to repeat the procedure until convergence. The programming of the IMA is done in a diagonal~\cite{hermesCor} or row-wise~\cite{narayanan2021fully} fashion, therefore takes considerably larger time (20$\times$ to 30$\times$ higher) than merely performing a parallel MVM. As a direct consequence, to map layers bigger than the cross-bar size we need to split the weights and the layer's execution on multiple IMAs, at the cost of additional area. 
However, having multiple IMAs allows reducing the occupancy of the generic memory of the system to store the weights, being them hosted by the cross-bar itself.

\begin{figure}[t]
    \centering
    \includegraphics[width=0.95\linewidth]{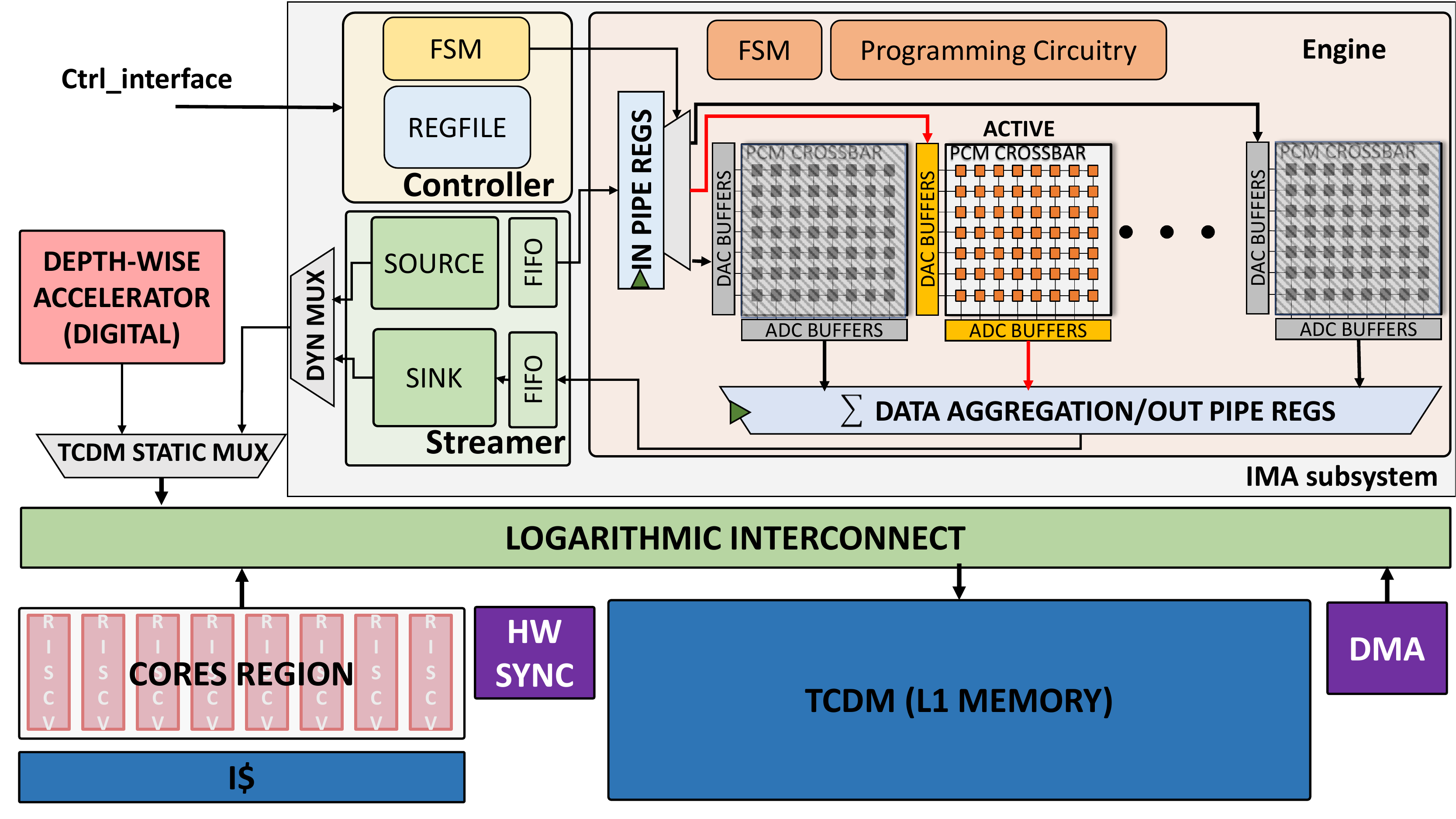}
    \caption{Overview of the scaled-up heterogeneous architecture. Only one IMC cross-bar can be active at a time. }
    \label{fig:discussion_archi}
    \vspace{-2mm}
\end{figure}

For this analysis, we integrate the IMC cross-bars into a single heterogeneous cluster of the same type presented in the previous sections. Specifically, multiple cross-bars are integrated into the same IMA sub-system, fully sharing the same data and control interface. They can be activated one at a time through a static multiplexing scheme. One multiplexer collects the data interfaces of the IMAs and redirects them into a 128-bit wide bus connected to the logarithmic interconnect of the cluster. However, they can all be programmed before the start of the computation, since we assume to replicate the configuration registers (mapped in different portions of the cluster memory map). Fig.~\ref{fig:discussion_archi} shows an overview of the architecture.


We adopt a sequential execution model for the layer-to-layer inference of the network, with the additional condition that all the input activations reside in the L1 memory of the cluster. In our analysis, we do not directly consider the overhead in terms of time and energy to access activation data from on-chip memory hierarchies. Double buffering and activation data tiling have been shown to be effective at hiding the time overhead\cite{burrello2020dory} and minimizing the energy one~\cite{rossi2021vega} in such cases, and we expect this effect to hold also in the case we analyze here.
%
In the case of the considered MobileNetV2 we map only the point-wise layers on the IMA cross-bars, while the depth-wise ones are executed on the digital accelerator. As reported in Sec.~\ref{sec:The_Bottleneck_Layer}, this solution leads to the highest performance and efficiency of the system.

\begin{figure*}[!t]
    \centering
    \includegraphics[width=0.95\linewidth]{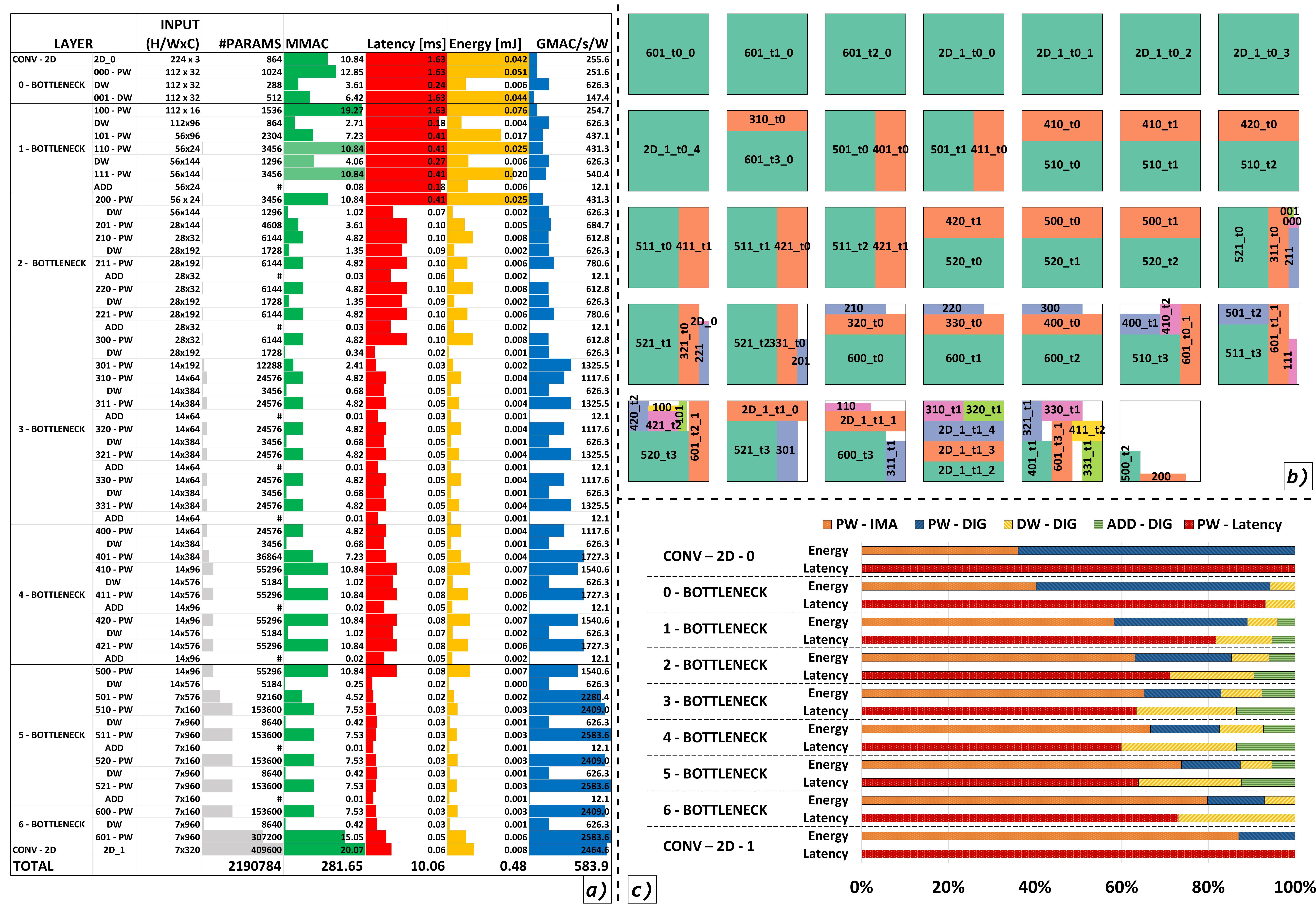}
    \caption{End-to-end execution of MobileNetV2 on the scaled-up heterogeneous cluster. (a) shows the parameters of each layer, the execution latency (ms), energy (mJ) and efficiency (GMAC/s/W). (b) Tiling algorithm of the layers on the 34 IMAs. (c) Latency and energy breakdown of \textit{Bottleneck} layers. }
    \label{fig:abl_study}
    \vspace{-2mm}
\end{figure*}
Only the first layers of the MobileNetV2 fit a single 256$\times$256 cross-bar, while the others (starting from the \textit{Bottleneck 3}) require to be split over multiple IMA tiles.
Therefore, we develop a \textsc{Tile\&Pack} strategy, outlined in Alg.~\ref{alg:tile_pack}, to tile all layers and pack their contributions in the smallest number of IMAs.
Tiling splits a layer over multiple IMAs only when it does not fit the size of the cross-bar; we do not allow tiling to fill unfilled IMA locations, aiming at the highest utilization area of the cross-bar on a per-tile basis.
Packing is based on the \textit{Maximal Rectangles Best Short Side Fit} bin fitting algorithm\footnote{We employ the open-source \textit{rectpack} Python library, available at \url{https://github.com/secnot/rectpack}~\cite{jylanki2010thousand}, to implement the \textsc{BinBestFit} and \textsc{MaxRectsBSSF} functions.}.
Fig.~\ref{fig:abl_study}(b) shows the result of the application of the \textsc{Tile\&Pack} algorithm to the weights of MobileNetV2.  From this analysis, we conclude that to map all the \textit{Bottleneck} layers of the MobileNetV2 we need 34 IMA cross-bars. As can be seen in Fig.~\ref{fig:abl_study}(b), the \textsc{Tile\&Pack} algorithm achieves 100\% of utilization of the cross-bar cells on most IMA cross-bars, with only the final one showing a utilization below 84\%.

The system with 34 IMAs would require a minimum area of $\sim$30 $mm^2$, since the area of the single IMA is 0.83 $mm^2$. Despite this might represent a drawback, it is worth noticing that weights need anyway to be stored into a non-volatile memory inside or outside the system, such as a Flash. In principle, the non-volatility of PCM-based IMAs allows eliminating this Flash memory from the system.

Each layer or layer tile considered in this study is benchmarked in terms of execution latency and energy individually, on the heterogeneous system analyzed in Sec.~\ref{sec:The_Bottleneck_Layer} (which incorporates only one IMA). We argue that, for this study which aims to be a guideline for further digital/analog systems explorations, this is a good approximation, since the benchmarked results include input/output fetch/storage from/to the L1 memory of the system and the instructions of the cores to configure and start the execution of the accelerators (this reasoning holds for point-wise, depth-wise and residual connection layers).

%
The results are shown in Fig.~\ref{fig:abl_study}(a), whereas in Fig.~\ref{fig:abl_study}(c) we report the energy and the latency breakdown (among the several hardware blocks involved in the computation) of the conv-2d and \textit{Bottleneck} layers. We notice higher execution latency and lower efficiency for point-wise layers with fewer parameters that operate on larger inputs -- typically, the ones from layers appearing early on in the network. In these cases, the major part of the energy is spent in digital logic, as these layers require more input and output streams to move the activations to be processed. The most efficient layers are the last two, where the IMA is utilized best, showing a peak of efficiency higher than 5 TOPS/W. Overall, the proposed architecture executes the end-to-end inference in 10.1ms of latency, consuming 482 $\mu J$.

%% file: text/07_SOA_Comparison_v2.tex
\input{text/Soa_table_v2}
\section{Comparison with the State-of-the-Art}
\label{sec:comparison_with_soa}

Tab.~\ref{tab:SOA_TABLE} reports the comparison of our scaled-up system with fully digital and mixed-signal state-of-the-art solutions.
The solution we propose is superior compared to the others, as it provides full hardware support for a wide range of workloads both in analog and digital domains, enabling \textit{de facto} efficient end-to-end execution of complex neural network models such as the MobileNetV2. Compared to Vega~\cite{rossi2021vega}, which is an architecture based on the same RISC-V cluster without integrating analog IMC cores nor dedicated accelerators for the depth-wise, we show 10$\times$ and 2.5$\times$ improvements in terms of inference latency and energy, respectively, when considering the end-to-end inference of the MobileNetV2. 

We compare favorably also with~\cite{jia2020programmable}, which consists of a tiny RISC-V core and a charge-based IMC array integrated into the system through a loosely-coupled scheme. In theory, the presence of a programmable core potentially enables the execution of a reasonably sized network such as MobileNetV2. However, the only processing model possible on this architecture is to offload the point-wise layers to the IMC array and the depth-wise and residual layers to the tiny RISC-V processor, which is not capable of performing compute-intensive functions with a reasonable performance level.
This would create a major performance bottleneck for the heterogeneous workload. For these reasons, our solution shows at least two orders of magnitude improvement on the end-to-end execution of the DNN. Despite the higher area of our system that might represent a drawback, it is worth noticing that the charge-based IMA integrated into~\cite{jia2020programmable} requires extending the architecture with a Flash memory to store the weights of the DNN (with non-negligible area costs). In our architecture, weights can be stored directly on the non-volatile IMAs, without having to consider an external Flash.

The system presented in~\cite{zhou2021analognets} consists of a PCM-based IMC array extended with digital logic that performs only activation and pooling operations, while a small SRAM memory acts as a layer-to-layer intermediate buffer. The higher peak performance and efficiency on MVMs they show compared to us is due to the bigger array size they used (1024$\times$512 compared to 256$\times$256), being the in-memory macro based on the same prototype~\cite{hermesCor}, while a loss as little as 10\% of the peak is attributable to the integration of the IMA into a complex system like the one we propose, as we show in Sec.~\ref{sec:bottlenecs_res}. However, the architecture in~\cite{zhou2021analognets} is too limited to execute the end-to-end inference of the MobileNetV2 for two main reasons: first, a single IMC array can not host all the layers weights of the MobileNet; second, there are no programmable cores to handle residual connections and control operations. Despite a more complex data-path compared to~\cite{zhou2021analognets}, including a cluster of 4$\times$4 computing in memory units and a network-on-chip for communication which delivers outstanding performance and efficiency on MVMs, also the architecture shown in~\cite{new_verma} is not viable to map heterogeneous workloads such as the MobileNetV2, due to the absence of a programmable processor.

\begin{figure}
    \centering
    \includegraphics[width=0.75\linewidth]{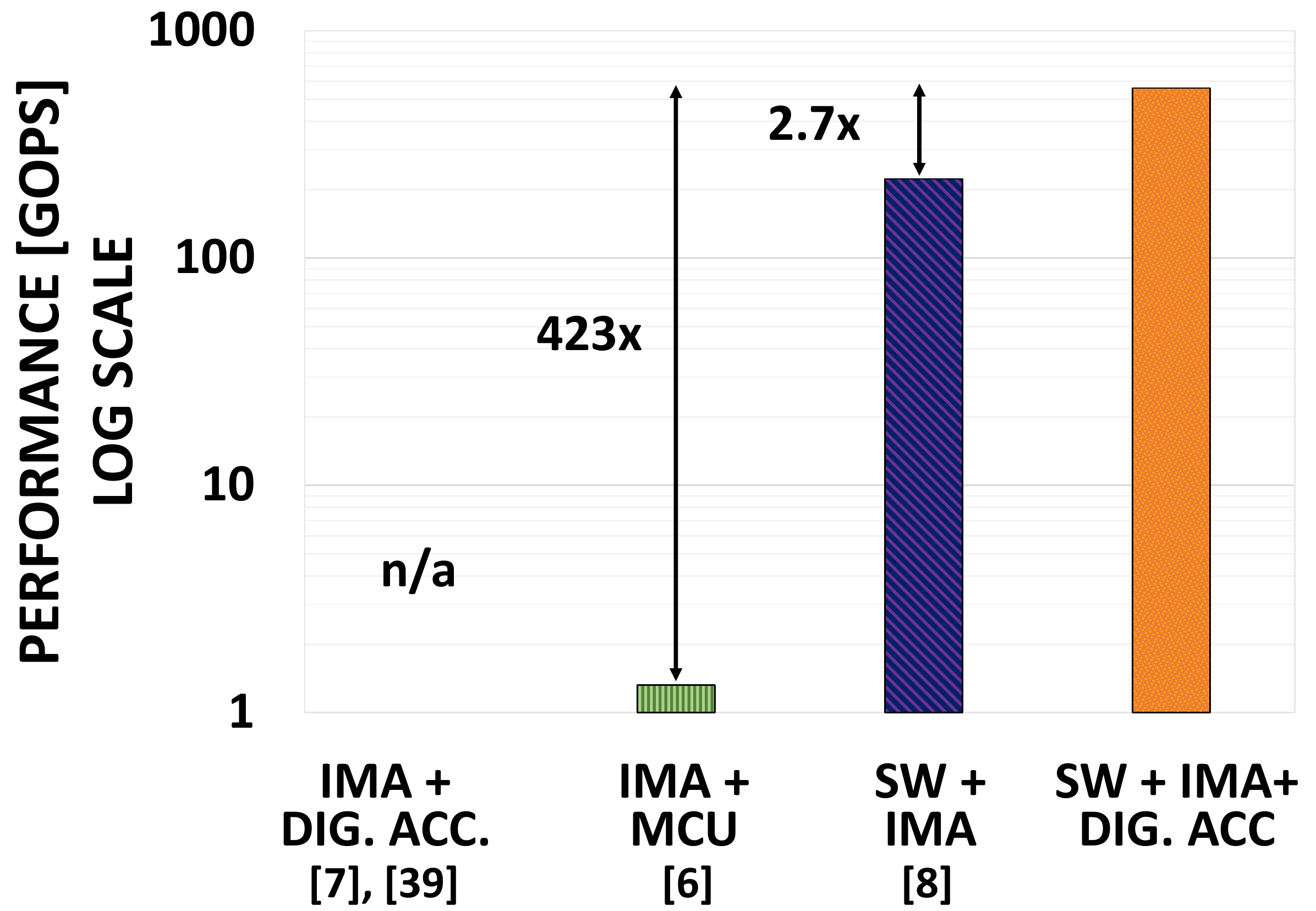}
    \caption{Performance of the MobileNetV2, on four IMC-based computing models. On the \textsc{ima+asic} it is not possible to deploy the network model, due to architectural limitations.}
    \label{fig:final_plot}
    \vspace{-2mm}
\end{figure}

Finally, to better highlight the contribution of this work, we abstract the specific System-on-Chip implementations described in Tab.~\ref{tab:SOA_TABLE} to four categories representative of the state-of-the-art, as shown in Fig.~\ref{fig:final_plot}. We can highlight four different processing models: \textit{i)} analog cores extended with fixed-function digital logic~\cite{zhou2021analognets, new_verma} (\textsc{ima+ dig. acc.}), \textit{ii)} analog cores controlled by simple MCU-subsystems~\cite{jia2020programmable} (\textsc{ima+ mcu}), \textit{iii)} IMAs integrated into tightly-coupled clusters of programmable processors~\cite{ottavi2021end} (\textsc{sw+ ima}), and \textit{iv)} the paradigm proposed in this work, where we exploit heterogeneity both in terms of analog and digital computing and in terms of programmable cores and lightweight tightly coupled digital acceleration (\textsc{sw+ima+ dig. acc.}).

Fig.~\ref{fig:final_plot} shows the results of the exploration, highlighting that for the MobileNetV2 workload, the computational model proposed in this work delivers significantly better performance compared to all the models exploiting programmable processor to sustain flexibility bottlenecks of IMC arrays. On the other hand, architectures only mixing specialized digital hardware with AIMC can only deal with DNN models for which they are designed, not being able to adapt to different models for which they were not intended before fabrication.

We argue that this concept, only demonstrated for MobileNetV2 DNN in Fig.~\ref{fig:final_plot} can be easily extended to more complex computer vision pipelines in the embedded domain, where AI workloads are often coupled to more traditional linear algebra algorithms such as Principal Component Analysis (PCA), Fast Fourier Transform, Filtering Functions or Inverse Kinematics~\cite{aristidou2018inverse}. We believe that the approach proposed in this work is a viable way to tackle the performance and flexibility requirements of rapidly evolving modern computer vision pipelines.

%% file: text/Soa_table_v2.tex
\begin{table}[t]
\centering
\caption{Comparison with the State-of-the-Art.}
\begin{tabular}{lccccc}
\cline{2-6}
\multicolumn{1}{l|}{}                                                                              & \multicolumn{1}{c|}{\textbf{\cite{rossi2021vega}}}                                                        & \multicolumn{1}{c|}{\textbf{\cite{zhou2021analognets}}}                                                       & \multicolumn{1}{c|}{\textbf{\cite{new_verma}}}                                                          & \multicolumn{1}{c|}{\textbf{\cite{jia2020programmable}}}                                                          & \multicolumn{1}{c|}{\textbf{\begin{tabular}[c]{@{}c@{}}This \\ Work\end{tabular}}}    \\ \hline
\multicolumn{1}{|l|}{\textbf{\begin{tabular}[c]{@{}l@{}}Tech. \\ node\end{tabular}}}               & \multicolumn{1}{c|}{22nm}                                                              & \multicolumn{1}{c|}{14nm}                                                             & \multicolumn{1}{c|}{16 nm}                                                                & \multicolumn{1}{c|}{65nm}                                                                 & \multicolumn{1}{c|}{22nm}                                                             \\ \hline
\multicolumn{1}{|l|}{\textbf{\begin{tabular}[c]{@{}l@{}}Area\\ {[}mm2{]}\end{tabular}}}            & \multicolumn{1}{c|}{12}                                                                & \multicolumn{1}{c|}{3.2}                                                              & \multicolumn{1}{c|}{25}                                                                   & \multicolumn{1}{c|}{13.5}                                                                 & \multicolumn{1}{c|}{$\sim$30}                                                         \\ \hline
\multicolumn{1}{|l|}{\textbf{\begin{tabular}[c]{@{}l@{}}Cores\\ (ISA)\end{tabular}}}               & \multicolumn{1}{c|}{\begin{tabular}[c]{@{}c@{}}9x\\  RV32\\ IMCF\\ Xpulp\end{tabular}} & \multicolumn{1}{c|}{None}                                                             & \multicolumn{1}{c|}{None}                                                                 & \multicolumn{1}{c|}{\begin{tabular}[c]{@{}c@{}}1 \\ RV32\\ IMC\end{tabular}}              & \multicolumn{1}{c|}{\begin{tabular}[c]{@{}c@{}}8x \\ RV32\\ IMC\\ Xpulp\end{tabular}} \\ \hline
\multicolumn{1}{|l|}{\textbf{\begin{tabular}[c]{@{}l@{}}Analog\\ IMC\end{tabular}}}                & \multicolumn{1}{c|}{None}                                                              & \multicolumn{1}{c|}{\begin{tabular}[c]{@{}c@{}}1x\\ PCM\end{tabular}}                 & \multicolumn{1}{c|}{\begin{tabular}[c]{@{}c@{}}16x\\ charge\end{tabular}}                 & \multicolumn{1}{c|}{\begin{tabular}[c]{@{}c@{}}1x\\ charge\end{tabular}}                  & \multicolumn{1}{c|}{\begin{tabular}[c]{@{}c@{}}34x\\ PCM\end{tabular}}                \\ \hline
\multicolumn{1}{|l|}{\textbf{\begin{tabular}[c]{@{}l@{}}Array\\ Rows\end{tabular}}}                & \multicolumn{1}{c|}{None}                                                              & \multicolumn{1}{c|}{1024}                                                             & \multicolumn{1}{c|}{1152}                                                                 & \multicolumn{1}{c|}{2304}                                                                 & \multicolumn{1}{c|}{256}                                                              \\ \hline
\multicolumn{1}{|l|}{\textbf{\begin{tabular}[c]{@{}l@{}}Array\\ Columns\end{tabular}}}             & \multicolumn{1}{c|}{None}                                                              & \multicolumn{1}{c|}{512}                                                              & \multicolumn{1}{c|}{256}                                                                  & \multicolumn{1}{c|}{256}                                                                  & \multicolumn{1}{c|}{256}                                                              \\ \hline
\multicolumn{1}{|l|}{\textbf{\begin{tabular}[c]{@{}l@{}}Digital\\ acc.\end{tabular}}}              & \multicolumn{1}{c|}{\begin{tabular}[c]{@{}c@{}}HWCE\\ (stand.\\ conv.)\end{tabular}}   & \multicolumn{1}{c|}{\begin{tabular}[c]{@{}c@{}}ReLu,\\ activ.,\\ im2col\end{tabular}} & \multicolumn{1}{c|}{\begin{tabular}[c]{@{}c@{}}Activ.,\\ scaling,\\ pooling\end{tabular}} & \multicolumn{1}{c|}{\begin{tabular}[c]{@{}c@{}}Activ.,\\ scaling,\\ pooling\end{tabular}} & \multicolumn{1}{c|}{\begin{tabular}[c]{@{}c@{}}Depth-\\ wise\end{tabular}}            \\ \hline
\multicolumn{1}{|l|}{\textbf{\begin{tabular}[c]{@{}l@{}}Peak \\ Perf. \\ {[}TOPS{]}\end{tabular}}} & \multicolumn{1}{c|}{\begin{tabular}[c]{@{}c@{}}0.032 \\ (ML 8b)\end{tabular}}          & \multicolumn{1}{c|}{\begin{tabular}[c]{@{}c@{}}2\\ (8b-4b)\end{tabular}}              & \multicolumn{1}{c|}{\begin{tabular}[c]{@{}c@{}}3\\ (8b-8b)\end{tabular}}                  & \multicolumn{1}{c|}{\begin{tabular}[c]{@{}c@{}}0.068 ${}^{1}$ \\ (8b-4b)\end{tabular}}             & \multicolumn{1}{c|}{\begin{tabular}[c]{@{}c@{}}0.958 \\ (8b-4b)\end{tabular}}         \\ \hline
\multicolumn{1}{|l|}{\textbf{\begin{tabular}[c]{@{}l@{}}Peak \\ Eff.\\ {[}TOPS/W{]}\end{tabular}}} & \multicolumn{1}{c|}{\begin{tabular}[c]{@{}c@{}}0.61 \\ (8b-8b)\end{tabular}}           & \multicolumn{1}{c|}{\begin{tabular}[c]{@{}c@{}}13.5\\ (8-4b)\end{tabular}}            & \multicolumn{1}{c|}{\begin{tabular}[c]{@{}c@{}}30\\ (8b-8b)\end{tabular}}                 & \multicolumn{1}{c|}{\begin{tabular}[c]{@{}c@{}}12.5 ${}^{1}$\\ (8b-4b)\end{tabular}}              & \multicolumn{1}{c|}{\begin{tabular}[c]{@{}c@{}}6.39 \\ (8b-4b)\end{tabular}}          \\ \hline
                                                                                                   & \multicolumn{1}{l}{}                                                                   & \multicolumn{1}{l}{}                                                                  & \multicolumn{1}{l}{}                                                                      & \multicolumn{1}{l}{}                                                                      & \multicolumn{1}{l}{}                                                                  \\
\multicolumn{6}{c}{\textbf{MobileNetV2 inference}}                                                                                                                                                                                                                                                                                                                                                                                                                                                                                                                  \\
                                                                                                   & \multicolumn{1}{l}{}                                                                   & \multicolumn{1}{l}{}                                                                  & \multicolumn{1}{l}{}                                                                      & \multicolumn{1}{l}{}                                                                      & \multicolumn{1}{l}{}                                                                  \\ \hline
\multicolumn{1}{|l|}{\textbf{\begin{tabular}[c]{@{}l@{}}Perf.\\ {[}inf./s{]}\end{tabular}}}        & \multicolumn{1}{c|}{10}                                                                & \multicolumn{1}{c|}{n/a}                                                              & \multicolumn{1}{c|}{n/a}                                                                  & \multicolumn{1}{c|}{0.23${}^{2}$}                                                                 & \multicolumn{1}{c|}{99}                                                               \\ \hline
\multicolumn{1}{|l|}{\textbf{\begin{tabular}[c]{@{}l@{}}Energy\\ {[}mJ{]}\end{tabular}}}           & \multicolumn{1}{c|}{1.19}                                                              & \multicolumn{1}{c|}{n/a}                                                              & \multicolumn{1}{c|}{n/a}                                                                  & \multicolumn{1}{c|}{n/a}                                                                  & \multicolumn{1}{c|}{0.482}                                                            \\ \hline
\end{tabular}
\label{tab:SOA_TABLE}
\begin{tablenotes}
    \item\footnotemark[1]{} Scaled from 1b-1b MVMs performance as explained in~\cite{jia2020programmable}.
    \item\footnotemark[2]{} Point-wise latency estimated from the peak performance on 8-bit$\times$4-bit MVMs. Latency of 8-bit$\times$8-bit depth-wise conv. estimated from our benchmarking results on the cluster's cores, scaled considering that: due to improved ISA, our core is $\sim$10$\times$ faster on a per-core basis~\cite{gautschi2017near}; additional $\sim$7$\times$ improvement factor due to the cluster parallelism~\cite{garofalo2020pulp}.
\end{tablenotes}
\vspace{-4mm}
\end{table}

%% file: text/08_conclusion.tex
\section{Conclusion}
\label{sec:conclusion}
Analog in-memory computing (IMC) promises outstanding improvements in energy efficiency on MVM operations. 
However, to target practical IoT applications IMC arrays must be enclosed in programmable heterogeneous systems, introducing new system-level challenges.

In this work, we explore these challenges by presenting a full implementation of a heterogeneous tightly-coupled clustered architecture integrating 8 RISC-V processors, a non-volatile PCM-based IMC accelerator, and a depth-wise digital accelerator, targeting the GlobalFoundries 22nm FDX technology.
Benchmarked on a highly heterogeneous workload such as the \textit{Bottleneck} layer, our solution overcomes software execution of the layer by 11.5$\times$ and 9.5$\times$ in terms of performance and energy efficiency. 
%
%
We scaled up our system to investigate the challenges and the resources of enabling end-to-end inference of real-world DNNs such as the MobileNetV2, demonstrating execution 10$\times$ faster within 2.5$\times$ lower energy than fully digital solutions and more than two orders of magnitude faster than existing state-of-the-art analog/digital heterogeneous solutions.
%


%% file: bio_text/authors_biography.tex
\begin{IEEEbiography}[{\includegraphics[width=1in,height=1.25in,clip,keepaspectratio]{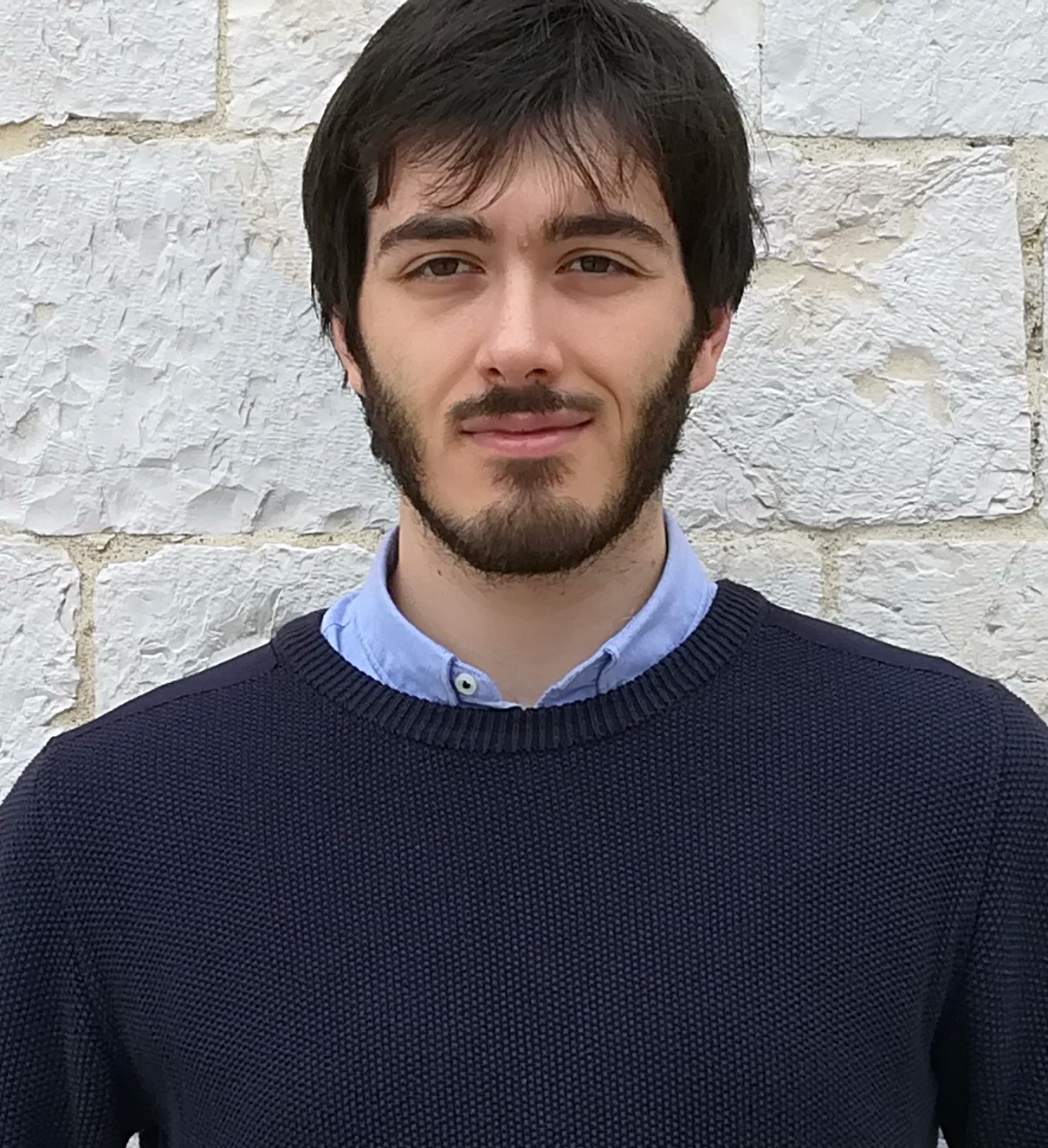}}]{Angelo Garofalo} received the B.Sc and M.Sc. degree in electronic engineering from the University of Bologna, Italy, in 2016 and 2018 respectively. He is currently working toward his Ph.D. degree at DEI, University of Bologna, Italy. His main research topic is Hardware-Software design of ultra-low power multiprocessor systems on chip for edge AI. His research interests include Quantized Neural Networks, Hardware efficient Machine Learning, in-memory computing heterogeneous architectures and fully-programmable embedded architectures.
\end{IEEEbiography}

\begin{IEEEbiography}[{\includegraphics[width=1in,height=1.25in,clip,keepaspectratio]{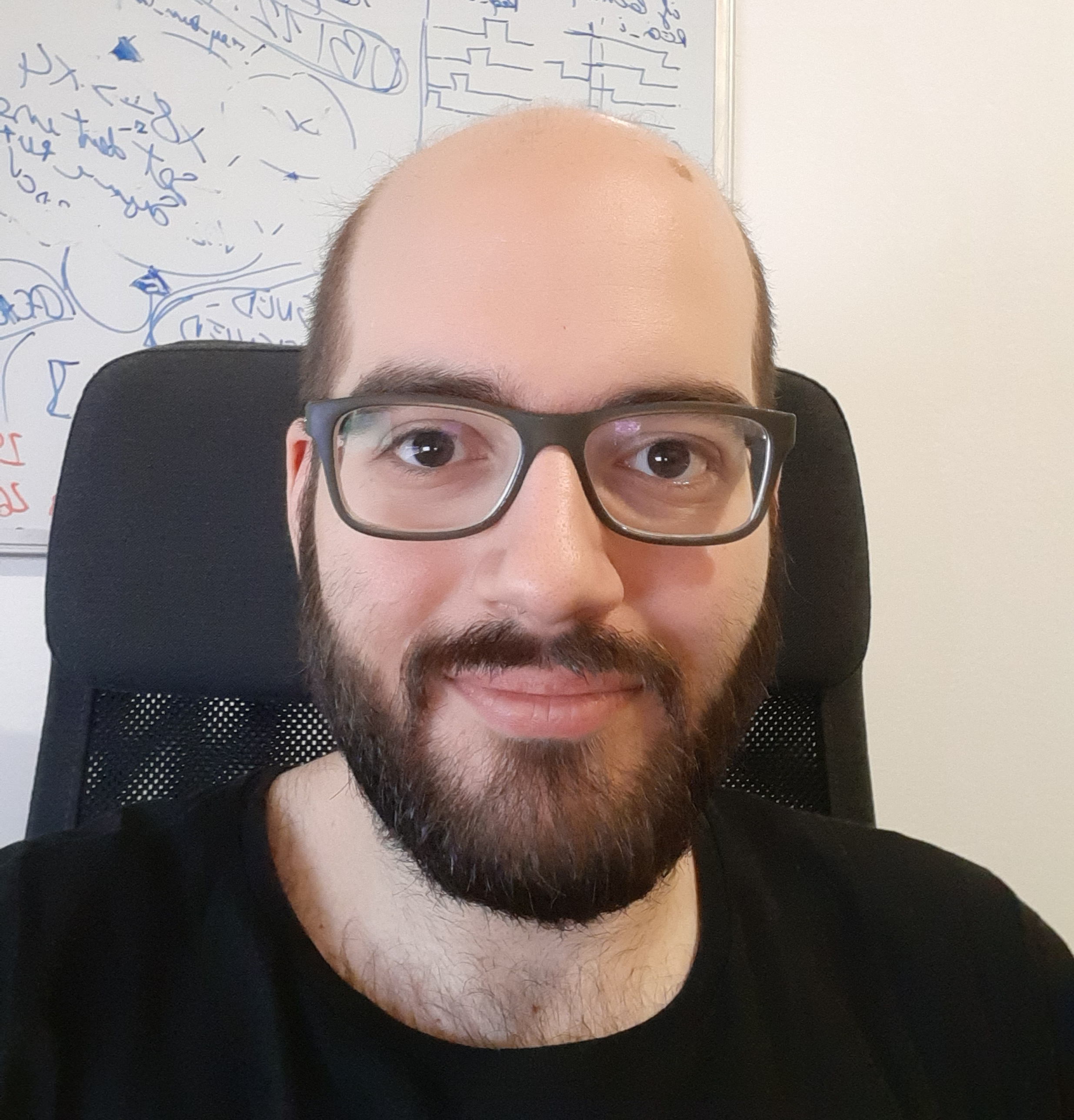}}]{Gianmarco Ottavi} recently started his Ph.D. in electronics Engineerig at the University of Bologna, Italy. After recieving his M.Sc. degree in 2019 he worked as a research Fellow for two years in the department of Electrical, Electronic and Information Engineering (DEI) in Bologna. Its field of research focused on hardware design for efficient inference in low-power systems where he developed specialized ISA extension for RISC-V and system-level implementation of In-memory computing accelerators.
\end{IEEEbiography}

\begin{IEEEbiography}[{\includegraphics[width=1in,height=1.25in,clip,keepaspectratio]{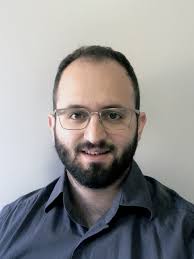}}]{Francesco Conti} received the Ph.D. degree in electronic engineering from the University of Bologna, Italy, in 2016. He is currently an Assistant Professor in the DEI Department of the University of Bologna. From 2016 to 2020, he held a research grant in the DEI department of University of Bologna and a position as postdoctoral researcher at the Integrated Systems Laboratory of ETH Zurich in the Digital Systems group. His research focuses on the development of deep learning based intelligence on top of ultra-low power, ultra-energy efficient programmable Systems-on-Chip. His research work has resulted in more than 40 publications in international conferences and journals and has been awarded several times, including the 2020 IEEE TCAS-I Darlington Best Paper Award.
\end{IEEEbiography}

\begin{IEEEbiography}[{\includegraphics[width=1in,height=1.25in,clip,keepaspectratio]{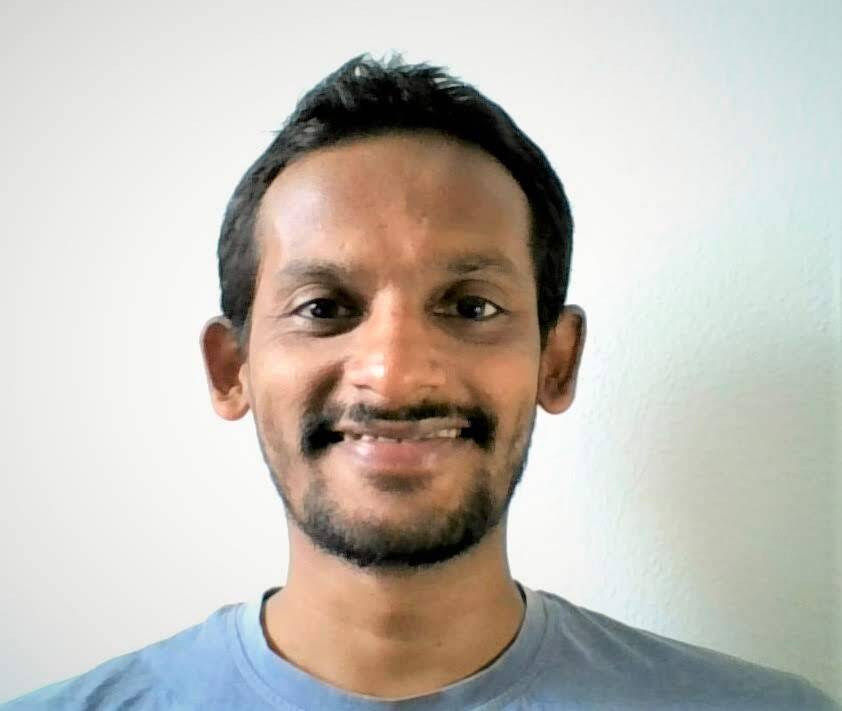}}]{Geethan Karunaratne} received his B.Sc. degree in Electronic and Telecommunication Engineering from University of Moratuwa, Sri Lanka in 2014, and the M.Sc. degree in Information Technology and Electrical Engineering from ETH Zurich in 2018. He joined IBM Research – Zurich in 2018, where he is currently a member of In-memory computing group. Geethan is working towards his PhD degree at ETH Zurich. His main research interests are in-memory computing and brain-inspired computing.
\end{IEEEbiography}

\begin{IEEEbiography}[{\includegraphics[width=1in,height=1.25in,clip,keepaspectratio]{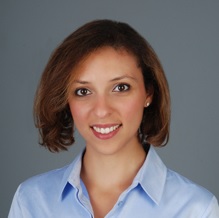}}]{Irem Boybat} (Member, IEEE) received her BSc degree in Electronics Engineering from Sabanci University, Turkey (2013), and MSc and PhD degrees in Electrical Engineering from Ecole Polytechnique Federale de Lausanne (EPFL), Switzerland (2015 and 2020, respectively). Since 2020, she is a postdoctoral researcher in the In-memory computing group of IBM Research Europe, Switzerland. Her research interests include in-memory computing for AI systems, neuromorphic computing and emerging resistive memory. She was a co-recipient of the 2018 IBM Pat Goldberg Memorial Best Paper Award, 2018 IBM Research Division Award on neuromorphic computing using phase-change memory devices and 2020 EPFL PhD Thesis Distinction in Electrical Engineering.
\end{IEEEbiography}

\begin{IEEEbiography}[{\includegraphics[width=1in,height=1.25in,clip,keepaspectratio]{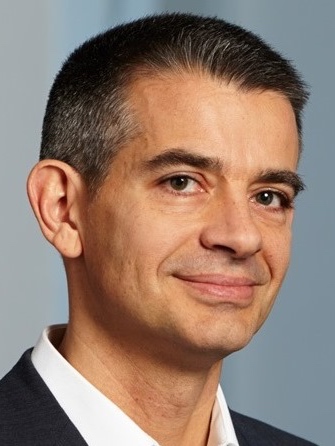}}]{Luca Benini} holds the chair of digital Circuits and systems at ETHZ and is Full Professor at the Università di Bologna. He received a PhD from Stanford University. Dr. Benini's research interests are in energy-efficient parallel computing systems, smart sensing micro-systems and machine learning hardware. He has published more than 1000 peer-reviewed papers and five books. He is a Fellow of the IEEE, of the ACM and a member of the Academia Europaea. He received the IEEE Mac Van Valkenburg award in 2016 and the ACM/IEEE A. Richard Newton Award in 2020.
\end{IEEEbiography}
\vspace{-160mm}
\begin{IEEEbiography}[{\includegraphics[width=1in,height=1.25in,clip,keepaspectratio]{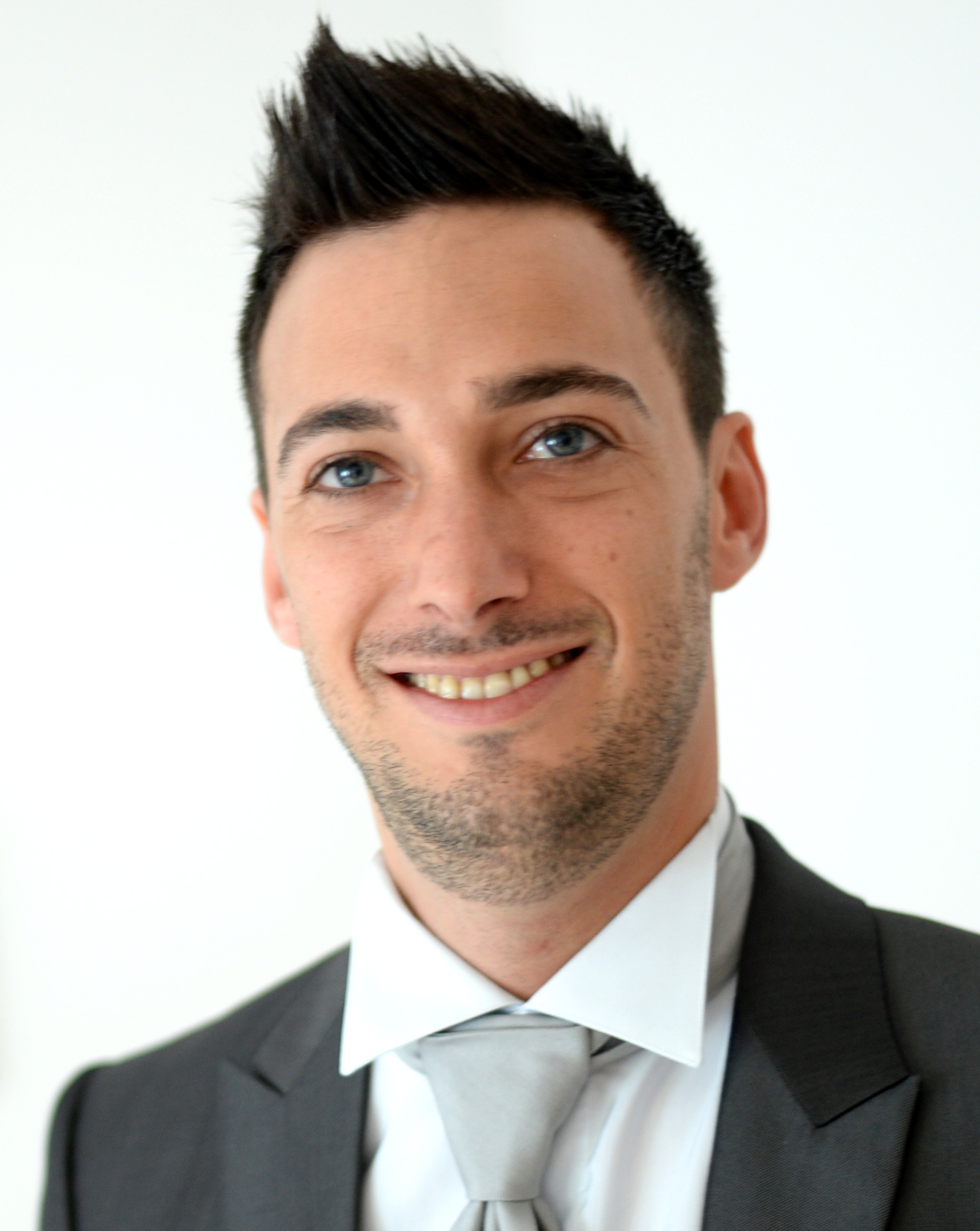}}]{Davide Rossi} received the Ph.D. degree from the University of Bologna, Bologna, Italy, in 2012. He has been a Post-Doctoral Researcher with the Department of Electrical, Electronic and Information Engineering “Guglielmo Marconi,” University of Bologna, since 2015, where he is currently an Assistant Professor. His research interests focus on energy-efficient digital architectures. In this field, he has published more than 100 papers in international peer-reviewed conferences and journals. He is recipient of Donald O. Pederson Best Paper Award 2018, 2020 IEEE TCAS Darlington Best Paper Award, 2020 IEEE TVLSI Prize Paper Award.
\end{IEEEbiography}